\documentclass[%
reprint,
 amsmath,amssymb,
 aps,
 superscriptaddress,
]{revtex4-2}

\usepackage{graphicx}
\usepackage{dcolumn}
\usepackage{bm}
\usepackage{hyperref}
\usepackage{longtable}

\usepackage[dvipsnames]{xcolor} 

\usepackage[mathlines]{lineno}

\begin{document}

\preprint{APS/123-QED}

\title{Probing the timescale dependency of local and global variations in surface air temperature from climate simulations and reconstructions of the last millennia
}

\author{Beatrice Ellerhoff}%
\email{beatrice.ellerhoff@iup.uni-heidelberg.de}
\affiliation{Institute of Environmental Physics, Ruprecht-Karls-Universität Heidelberg, INF 229, 69120 Heidelberg, Germany 
}%
\author{Kira Rehfeld}
\affiliation{Institute of Environmental Physics, Ruprecht-Karls-Universität Heidelberg, INF 229, 69120 Heidelberg, Germany
}
\affiliation{now at: Geo- und Umweltforschungszentrum (GUZ), Universität Tübingen,
Schnarrenbergstr. 94-96, 72076 Tübingen, Germany}

\date{Published 27 December 2021 in Physical Review E}

\begin{abstract}
Earth's climate can be understood as a dynamical system that changes due to external forcing and internal couplings. Essential climate variables, such as surface air temperature, describe this dynamics. Our current interglacial, the Holocene (11,700 yr ago to today), has been characterized by small variations in global mean temperature prior to anthropogenic warming. However, the mechanisms and spatiotemporal patterns of fluctuations around this mean, called temperature variability, are poorly understood despite their socio-economic relevance for climate change mitigation and adaptation. 
Here, we examine discrepancies between temperature variability from model simulations and paleoclimate reconstructions by categorizing the scaling behavior of local and global surface air temperature on the timescale of years to centuries. To this end, we contrast power spectral densities (PSD) and their power-law scaling using simulated and observation-based temperature series of the last 6000 yr. We further introduce the spectral gain to disentangle the externally forced and internally generated variability as a function of timescale. It is based on our estimate of the joint PSD of radiative forcing, which exhibits a scale break around the period of 7 yr. We find that local temperature series from paleoclimate reconstructions show a different scaling behavior than simulated ones, with a tendency towards stronger persistence (i.e., correlation between successive values within a time series) on periods of 10 to 200 yr. Conversely, the PSD and spectral gain of global mean temperature are consistent across data sets. Our results point to the limitation of climate models to fully represent local temperature statistics over decades to centuries. By highlighting the key characteristics of temperature variability, we pave a way to better constrain possible changes in temperature variability with global warming and assess future climate risks. 

\end{abstract}

\maketitle

\section{\label{sec:introduction}Introduction}

The variability of surface air temperature is present on all spatial and temporal scales, from synoptic and seasonal changes to long-term variations on periods of years to multi-millennia. On the one hand, it arises from internal processes, such as the El Niño-Southern Oscillation (ENSO) \cite{Bjerknes1966}. On the other hand, the temperature varies due to external forcing, such as the greenhouse effect \cite{Arrhenius1896, Fourier1824}. Understanding the internally generated and externally forced variability has been suggested to be at least as necessary for evaluating climate risks for society and ecosystems as projecting the global mean temperature \cite{Katz1992}. Available instrumental observations are limited to a small time span, leading to challenges in quantifying temperature variability. Paleoclimate reconstructions extend the characterization of temperature variability and can be compared to global circulation models (GCMs) \cite{Ghil2020, Franzke2020, Tierney2020}. However, discrepancies between model and paleoclimate data remain to be resolved, especially on the local level and on periods between years and centuries \cite{Laepple2014GRL, Laepple2014PNAS, Parsons2017, Ljungqvist2019, Buehler2020}.

Characterizing local temperature variability is crucial for predicting extremes \cite{Franzke2020}, not only to minimize short-term damage but also to design long-term strategies, including urban planning and food cultivation \cite{Anderson2019}. Variability of global temperature on periods above years is relevant to the understanding of long-term changes \cite{Crucifix2017} as well as climate sensitivity \cite{Rypdal2018}. Assessing the temporal correlation structure of temperature series by means of scaling behavior and persistence is particularly important for distinguishing externally forced trends from natural changes \cite{Franzke2017}. It could affect the confidence in future projections and attribution studies \cite{Barnett1999, Mirle2013}. Therefore, one of the main topics to be investigated here is the characteristics of local and global temperature variability on periods of years to centuries from model simulations and observation-based data of the last millennia. 

To determine how the variability of a temperature series is distributed with timescales $\tau$, we make use of the power spectral density (PSD) $S(\tau)$, known as ``spectrum”. It can be obtained from the Fourier transform of the autocorrelation function (see Appendix \ref{app:PowerSpectrum}) \cite{Wiener1930, Khintchine1934}. The spectrum was shown to often follow a power law 
\begin{equation}
    \label{eq:spectrum}
    S(\tau) \sim \tau^{\beta} \,,
\end{equation}
with spectral exponent $\beta$ and period $\tau$ \cite{Wunsch2003, Franzke2012, Huybers2006, Lovejoy2015, Nilsen2016, Fredriksen2017}, especially on decadal-to-centennial scales \cite{Fraedrich2003a, Rypdal2013, Zhu2019a}. We refer to this behavior \eqref{eq:spectrum} as temporal scaling since the temperature signal has no preferred timescale and is statistically similar across periods $\tau$. The exact determination of the start and end points of a scaling interval is not part of this study. 

Long-range memory stochastic processes are suitable to describe temperature signals with temporal scaling \cite{Nilsen2016, Lovejoy2015, Mandelbrot1968}. Among those, fractional Gaussian noise (fGn) is a stationary process and exhibits a spectral exponent $\beta \in \left(-1, 1\right)$ on sufficiently long periods (see Appendix \ref{app:Autocovariance}). Fractional Brownian motion (fBm) is a nonstationary process that shows $\beta \in \left(1, 3\right)$. The scaling exponent $\beta$ relates to the decay of the autocovariance function (see Appendix \ref{app:Autocovariance}) and indicates how strongly the values within a time series are correlated (or anticorrelated). It is therefore regarded as a measure of the strength of temporal persistence (or antipersistence) \cite{Malamud1999, Nilsen2016}.

\begin{figure}
\includegraphics[width=8.6cm]{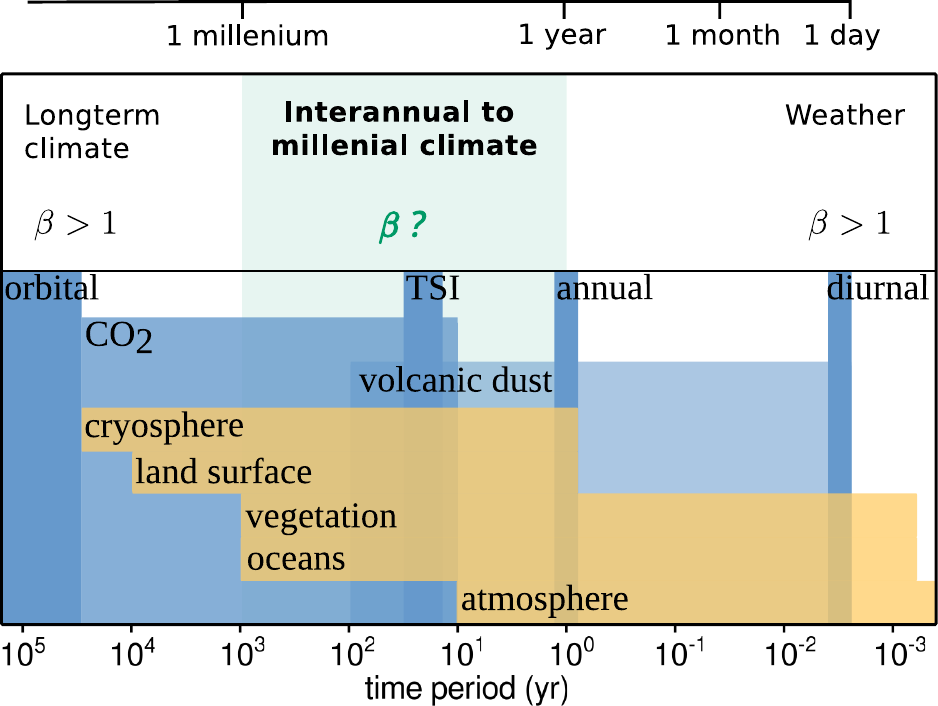}
\caption{\label{fig:climatesystem} Characteristic timescales relevant to surface air temperature (T) variability of climatic drivers (dark blue) and climate subsystems (yellow) \cite{Peixto1984, Rohling2012, Rohling2018}. The weather and long-term climate is characterized by $\beta >1$ for local and global mean temperature. 
On interannual to millennial timescales the statistical properties of temperature fluctuations remain to be determined, especially at the local scale. The TSI bar highlights the dominant variations in recent total solar irradiance observations.} 
\end{figure} 

Particular scaling behavior with $\beta \approx 2$ \cite{Lovejoy2015, Franzke2012} is typical for the weather regime (hours to weeks) and can be explained by atmospheric turbulence \cite{Zhang2019, Pelletier2002}. In the long-term climate, regional and global mean temperatures show similar behavior ($\beta > 1$) \cite{Lovejoy2015, Huybers2006, Zhu2019a} due to the presence of nonlinear processes, such as the temperature-albedo feedback \cite{Lohmann2018}. On timescales between years to millennia, the temperature is constantly influenced by the interaction of all climate subsystems and by volcanic, solar, as well as CO$_2$ forcing (Fig.~\ref{fig:climatesystem}). Estimates of the spatially-dependent scaling behavior of local temperature on these timescales differ \cite{Lovejoy2015, Fraedrich2003a}. On the global scale, many studies find $\beta \approx 1$ \cite{Rypdal2013, Zhu2019a}. However, Lovejoy \textit{et al.} has identified a change from the so-called macroweather regime ($\beta \approx 0.8 $ on periods of 10 days to 40 yr) to the climate regime ($\beta \approx 1.8$ on periods from 40 yr to 80 000 yr) \cite{Lovejoy2015}. 

In this manner, previous works find ambiguity in the interpretation of local and global temperature scaling and it remains to be determined whether simulations and reconstructions qualitatively agree in scaling behavior $\beta < 1$ or $\beta > 1$. The so-called ``$1/f-$noise'' ($\beta = 1$) corresponds to a process with power spectral density proportional to the period. For $\beta >1$, 
the relative contribution 
\begin{equation}
    \frac{\int_{f'/2}^{f'} S(f) \, \mathrm{d}f }{ \int_{f'}^{2f'} S(f) \,\mathrm{d}f} = \frac{1-2^{\beta-1}}{2^{1-\beta}-1} = 2^{\beta-1}
\end{equation}
to the variance is larger from slow timescales compared to faster ones for all frequency intervals $f'/2$ to $2f'$ within a scaling interval \cite{Lovejoy2019}. With increasing $\beta > 1$, the fBM is said to exhibit ``nonlinear pseudo-trends'' \cite{Mandelbrot1968} (see Appendix \ref{app:Autocovariance}). Thus, for understanding climate variability and for modeling purposes, the systematic estimate of the scaling exponent $\beta$ allows to assess the behavior of fluctuation levels across timescales \cite{Nilsen2016}. Moreover, the differentiation between forced and unforced changes poses a challenge to understanding temperature variability \cite{Marvel2016, Schurer2013}. Beyond the analysis of Haar fluctuations of a few forcing reconstructions \cite{Lovejoy2012, Lovejoy2016, RypdalRypdal2016}, spectral analysis of climatic drivers and their frequency-dependent linkage to the temperature response remains incomplete. 

We investigate the timescale dependency of local and global surface air temperature variability by analyzing power spectral densities from a few hours to a thousand years, thereby extending and improving on earlier work \cite{Mitchell1976a, Lovejoy2015, Huybers2006, Zhu2019a}. We use model simulations and observation-based data, which we introduce in Sec.~\ref{sec:data}. To estimate the PSD and determine its power-law scaling on periods of 10 to 200 yr, we use state-of-the-art methods described in Sec.~\ref{sec:methods}. This allows us to contrast regional and global spectra (Sec.~\ref{sec:gmstrmst}), spatial patterns (Sec.~\ref{sec:scalingmap2}), and the agreement of simulated and observation-based estimates (Sec.~\ref{sec:stat}). Along with that, we discuss the joint PSD from various radiative forcings, which allows us to calculate the spectral gain and study the externally forced variability in Sec.~\ref{sec:gain}. Based on our reconstruction of the PSD of surface air temperature for the last millennia, we evaluate the consistency of spectral characteristics across the data sets considered. In Sec.~\ref{sec:conclusion}, we elaborate on the stronger persistence of temperature on local than global level as well as remaining discrepancies. Finally, we discuss how our findings could help improve climate model simulations and understand Earth's climate dynamics.  

\section{\label{sec:data}Data} 
We investigate the timescale-dependent distribution of surface air temperature variability using model simulations, observation-based data, and radiative forcing reconstructions. The model simulations include ten transient runs from GCM experiments \cite{Henderson-Sellers1987}. 
The observation-based data consists of reanalysis data, instrumental measurements, and the paleoclimate reconstructions from the Past Global Changes 2k (PAGES2k) network \cite{Neukom2019}. We use 12 reconstructions of climatic drivers, including solar, volcanic, orbital, and CO$_2$ forcing. All temperature and radiative forcing signals are specific to the Mid- and Late-Holocene (the last 6000 yr), with a focus on the Common Era (0 to 2000 CE). The supplemental tables S1-S3 \cite{Supplementary} summarize their key specifications.

\subsection{Model simulations}
Each of the ten GCM runs considered features a transient, albeit different forcing and a comparable spatiotemporal resolution. The CESM-LME 1 \cite{Otto-Bliesner2016} and MPI-M LM \cite{Jungclaus2010} experiments serve as representative runs of the last millennium. We analyze them at two temporal resolutions (one month, six hours) to capture both the high- and low-frequency variability within our available computing capacities (see Fig. S7 \cite{Supplementary}). CESM 1 past 2k \cite{Zhong2018} is included as a slightly newer run for the Common Era. To cover the Mid-Holocene, we use simulations from the IPSL \cite{Braconnot2019} (denoted IPSL-p6k) and ECHAM5/MPI-OM  \cite{Fischer2011} (denoted ECH5/MPIOM-p6k) of the last 6000 yr. From the TraCE-21k \cite{Liu2009} simulation, we also consider only the last 6000 yr to retain comparability and to avoid potential biases due to significant shifts in the mean state of climate. The Mid-Holocene runs were temporally averaged to a bi-monthly resolution to reduce computational costs. To test for the influence of human-induced climate change on our results, we include the HadCM3 LM1 simulation \cite{Buehler2020}, covering the period from 850 to 1850 CE. Furthermore, we compare our results to the pre-industrial (PI) control runs from CESM-LME 1 and MPI-M LM, as well as the TraCE-21k-ORB run, which is solely forced by orbital changes.

\subsection{Observation-based data}

In addition to the simulations, we analyze the monthly resolved HadCRUT4 (Hadley Centre/Climatic Research Unit Temperature) instrumental records, ranging from 1850 to 2019 \cite{Morice2012}. However, most of the grid-box time series are not available as continuous measurements as required for spectral analysis. Therefore, we retain only those 104 grid boxes with coverage greater than 150 yr after interpolating gaps of up to two months. While the Northern Hemisphere is comparatively well covered up to 72.5$^\circ$N, only nine grid boxes remain for the Southern Hemisphere. Therefore, this selection comes at the expense of spatial resolution but offers a higher spectral resolution on longer timescales. To further explore the potential effect of these spatiotemporal constraints, we include the ERA5 (European Centre for Medium-Range Weather Forecasts Reanalysis 5th generation) temperature reanalysis for the years 1979 to 2019 \cite{Hersbach2020}. Along with CESM-LME 1 and MPI-M LM, we analyze the ERA5 data at both six-hourly and monthly resolution for the same reasons as mentioned earlier.

In addition to direct temperature observations and reanalysis, we analyze paleoclimate data. Paleoclimate records hold preserved biological, chemical, and physical tracers (``proxies'') of past climate. The number of temperature records from paleoclimate data with sub-centennial resolution is limited. Recent progress has been made by improved calibration and pseudo-proxy methods within the PAGES2k network \cite{Emile-Geay2017}. Therefore, we base our analysis on their newest global multiproxy database for temperature reconstructions of the Common Era \cite{Neukom2019}. It gathers 692 records from trees, ice, sediment, corals, speleothems, and documentary evidence with a resolution between weeks and centuries. The records are spread over 648 locations, including all continental regions and major ocean basins. 

For investigating the variability of global mean surface temperature, we use the seven spatially-weighted statistical reconstructions for the last 2000 yr provided by PAGES2k \cite{Neukom2019}. To estimate the mean of local spectra, we choose records from the PAGES2k database according to their resolution ($\leq 80 $yr), their number of data points ($\geq 20$), their coverage ($\geq 20$ yr), as well as their maximum hiatuses ($\leq$ 160 yr). To reliably deduce the scaling of the PSD from individual records, we select the records according to our scales of interest (\autoref{tab:timescales}), similar to \cite{Kantelhardt2011, Nilsen2016}. Ice core records were excluded from our analysis since they require additional consideration of  signal-to-noise ratios at the sub-centennial timescales \cite{Laepple2018, Casado2020}. 

\begin{table}
\caption{\label{tab:timescales} Requirements on irregularly sampled time series $x(t)$ for analyzing power-law scaling on timescales $\tau \in \left[\tau_1, \tau_2\right]$. We apply this scheme for $\tau_1=10$ and $\tau_2=200$ yr in Sec.~\ref{sec:scalingmap2} and \ref{sec:statistics}.} 
\begin{ruledtabular}
\center
\begin{tabular}{ll}
Parameter  & Value \\ \hline
Number of data points  ($N$) & $\geq$ 50  \\
Mean temporal resolution  ($\langle t_{i+1}$ - $t_i \rangle$) & $\leq \tau_1$ \\
Coverage ($t_N$ - $t_1$) &  $\geq 3 \tau_2$ \\
Length of hiatuses (max($t_{i+1}$ - $t_i$)) & $\leq 5 \tau_1$
\end{tabular}
\end{ruledtabular}
\end{table}

\subsection{Radiative forcing}

External forcing contributes significantly to temperature variability and is an essential part of reliable climate projections \cite{Schurer2013, Crowley2000, Hegerl2007}. We study its spectral properties using forcing reconstructions, widely implemented in GCM experiments and coordinated within the Palaeoclimate Model Intercomparison Project (PMIP3/PMIP4) \cite{Braconnot2012, Schmidt2012}. This includes five solar \cite{Delaygue2011, Steinhilber2009a, Wang2005, Muscheler2007, Vieira2010}, one CO$_2$ \cite{Schmidt2012} and two volcanic \cite{Crowley2000, Gao2008} forcing reconstructions as well as Berger’s numerical solution for orbital forcing \cite{Berger1978}. Furthermore, we calculate diurnal insolation changes from the hour angle of the sun \cite{Crucifix2016}. We also use a more recently published volcanic \cite{Toohey2017} and high-resolution solar forcing \cite{Frohlich2006} reconstruction as well as CO$_2$ measurements \cite{Keeling1976}. We neglect land-use forcing \cite{Pongratz2008} which is much lower in amplitude and variability than the other forcings considered here. 

All forcing reconstructions are rescaled to radiative forcing equivalents, which express their respective change in the Earth's radiation balance in Watts per square meter ($\mathrm{W m^{-2}}$). We apply the widely used formula $5.35\, \mathrm{ln}(\left[\mathrm{CO}_2\right] / 278 \mathrm{ppm}) \, \mathrm{W m^{-2}}$ to rescale CO$_2$ concentrations $\left[\mathrm{CO}_2\right]$, given in parts per million (ppm) \cite{Myhre1998}. The stratospheric aerosol optical depth (AOD) from volcanic eruptions is rescaled by $(-20)^{-1}$ $\mathrm{W m^{-2}}$/AOD \cite{Schmidt2011}, however, the optimal conversion factor is still a matter of debate \cite{IPCC2013}. Additional uncertainties arise from the wide spread of reconstructions for volcanic and solar forcing. To account for this and the choice of conversion factor, we simulate the joint PSD of radiative forcing by a Monte Carlo approach described in Appendix \ref{app:MonteCarlo}. Here, ``joint'' indicates that the PSD of radiative forcing is calculated by linear summation of the mean PSD from different types of climatic drivers, rescaled to their radiative forcing equivalents. 

\section{\label{sec:methods}Methods}

Spectral analysis is the primary tool used here for studying the timescale-dependent variability and scaling of temperature series. To minimize uncertainties in the spectral analysis of proxy records, we use state-of-the-art approaches for irregularly sampled time series \cite{Laepple2013}. Statistical estimators further test for the agreement between simulations and paleoclimate data. We apply linear response theory to derive the spectral gain and investigate the forced temperature response. 

\subsection{Spectral analysis}

Power spectral analysis requires the assumption that the underlying time series can be described as a weakly stationary, stochastic process with time-independent mean and autocovariance \cite{Chatfield2019}. We therefore linearly detrend all time series as it is standard for temperature analysis \cite{Fredriksen2016, Nilsen2016, Rehfeld2018, Laepple2014PNAS}. The agreement of the PSD from disjoint time intervals in Fig. S13 \cite{Supplementary} provides evidence that stationarity is sufficiently fulfilled. We use the multitaper method with three windows \cite{Percival1993, Yiou1996} and chi-square distributed uncertainties to compute the PSD. The two lowest frequencies were omitted to reduce biases of the multitaper method \cite{Huybers2006}. For visual purposes, we apply a logarithmic Gaussian smoothing filter of constant width (0.005 decibels) \cite{Kirchner2005}. Mean spectra were calculated by interpolation to the lowest resolution, binning into equally spaced log-frequency intervals, and taking the average with equal weights \cite{Huybers2006}. This requires the statistical independence of the averaged values \cite{RypdalRypdal2016}.  The spectral exponent $\beta$ is calculated by linear regression to the logarithm of \eqref{eq:spectrum} on periods between $\tau_1=10$ and $\tau_2=200$ yr after binning the PSD into equally spaced log-frequency intervals to more uniformly weight the estimate and avoid low-frequency biases \cite{Huybers2006, Nilsen2016, Zhu2019a, Ostvand2014}. In the case of seven proxy records with an insufficient resolution, the scaling is estimated on their corresponding spectral resolution, but always at least between 20 and 200 yr (Fig. S4 \cite{Supplementary}). The uncertainty of the spectral exponent, $\Delta \beta$, is given by the standard error of the linear regression model $\Delta \beta_{lm}$, except for irregularly temperature series.

\subsection{\label{sub:uncertainties} Uncertainties for irregular temperature series}

Spectral analysis of proxy records, which are typically not sampled in regular time steps, is more prone to errors than that of regular time series. We aim to minimize biases by accounting for the number of data points, temporal resolution, total coverage, and hiatuses' length when selecting the records (\autoref{tab:timescales}). We find that the mean temporal resolution of a proxy record approximates well the optimal interpolation time step. Nevertheless, the interpolation introduces uncertainties which are not captured by $\Delta \beta_{lm}$. Similar to Laepple \textit{et al.} \cite{Laepple2013}, we quantify this additional uncertainty $ \Delta \beta_{int}$ in four steps: (1) For each record with spectral exponent $\beta$, we simulate $N=100$ surrogate time series with annual resolution and a power-law scaling $\beta_n \approx \beta$ and $n \in [1,N]$. (2) We form the surrogate's block-average over the proxy record's irregular time steps and obtain $N$ surrogate time series at record resolution. (3) We interpolate the surrogate time series, calculate the multitaper spectrum, and extract the scaling exponent $\beta_{n, lm}$ from linear regression in the same way as for the proxy record (Fig. S8 \cite{Supplementary}). (4) We calculate the mean deviation $\Delta \beta_{int} = \frac{1}{N} \sum_{n=1}^{N} \vert \beta_{n, lm} - \beta_n \vert$ of the ensemble. The uncertainty of the individual fits $\Delta \beta_{n, lm}$ is negligible compared to the mean deviation $\Delta \beta_{int}$. We obtain the uncertainty of the record's spectral exponent from both, the uncertainty of the initial fit $\Delta \beta_{lm}$ and due to interpolation $\Delta \beta_{int}$ via quadratic summation: $\Delta \beta = \sqrt{ (\Delta \beta_{lm})^2 + (\Delta \beta_{int})^2 }$.

\subsection{\label{sec:stat}Statistical analysis of spectral exponents}
 
We quantify the agreement of simulated and reconstructed $\beta$-values using percent agreement, categorical agreement, and Kappa statistics. Beforehand, we extract the simulated temperature at the proxy record location by bilinear interpolation of neighboring grid boxes to achieve the best possible comparability between record and simulation. Percent agreement $p_0$ gives the percentage of locations at which the confidence range $\beta \pm \Delta \beta$ from simulation and reconstruction overlap. The agreement by category, here referred to as categorical agreement $p_c$, is calculated with the help of $\nu=0.32$, the mean uncertainty of $\beta$ from all proxy records considered. We then assign the three categories \textit{low} ($\beta < 1 - \nu $), \textit{high} ($1 + \nu \leq \beta $) and \textit{intermediate} ($1- \nu \leq \beta < 1 + \nu$) to the spectral exponent $\beta$. The \textit{intermediate} regime prevents incorrect assignment. To verify the reliability of categorical agreement, we calculate the Kappa statistics 
\begin{equation}
     \kappa= (p_c - p_e)/(1-p_e)
\end{equation}
with expected percent agreement $p_e$ by category \cite{Fleiss1973}. The latter can be obtained from $p_{e} = \frac{1}{N^2}\sum^{3}_{c=1} n_{c,m} n_{c,p}$ where $c$ is the category, $N$ the number of locations and $n$ the number of times that models ($m$) and proxy records ($p$) have predicted category $c$.  The $\kappa$-coefficient quantifies the reliability from no agreement beyond chance ($\kappa=0$) to full agreement ($\kappa=1$). Negative $\kappa$ indicates agreement that is beyond change, for example, due to systematic biases. 

\subsection{\label{sec:gain}Spectral gain}
We investigate how climatic drivers influence the global mean temperature at period $\tau$ by calculating the spectral gain 
\begin{equation}
\label{eq:gain}
     G^2(\tau) = \dfrac{S_T(\tau)}{S_F(\tau)}\,.
\end{equation}
Here, $S_T(\tau)$ is the PSD of the global mean temperature and  $S_F(\tau)$ the PSD of radiative forcing (see also Appendix \ref{app:Gain}). The gain requires the assumption that the global mean temperature can be well approximated as a linear function of the forcing \cite{Geoffroy2013, MacMynowski2011, Fredriksen2017} and that different types of radiative forcing add linearly \cite{Kirkevag2008, Shiogama2013, Meehl2004, Ramaswamy1997}. To this end, we focus on timescales between years and centuries when additivity is a valid assumption and nonlinearities in the global mean temperature are sufficiently small \cite{Lovejoy2016, RypdalRypdal2016}. The main practical problem that confronts us is that the gain might be subject to a sampling bias due to our data sets choice. Therefore, we perform a Monte Carlo simulation of the PSD of radiative forcing and the global mean temperature, as well as the spectral gain as described in Appendix \ref{app:MonteCarlo}.

\section{\label{sec:results} Results and discussion} 

\subsection{\label{sec:gmstrmst} Global mean and mean of local spectra}

\begin{figure*}[!htbp]
	\includegraphics[width=17.2cm]{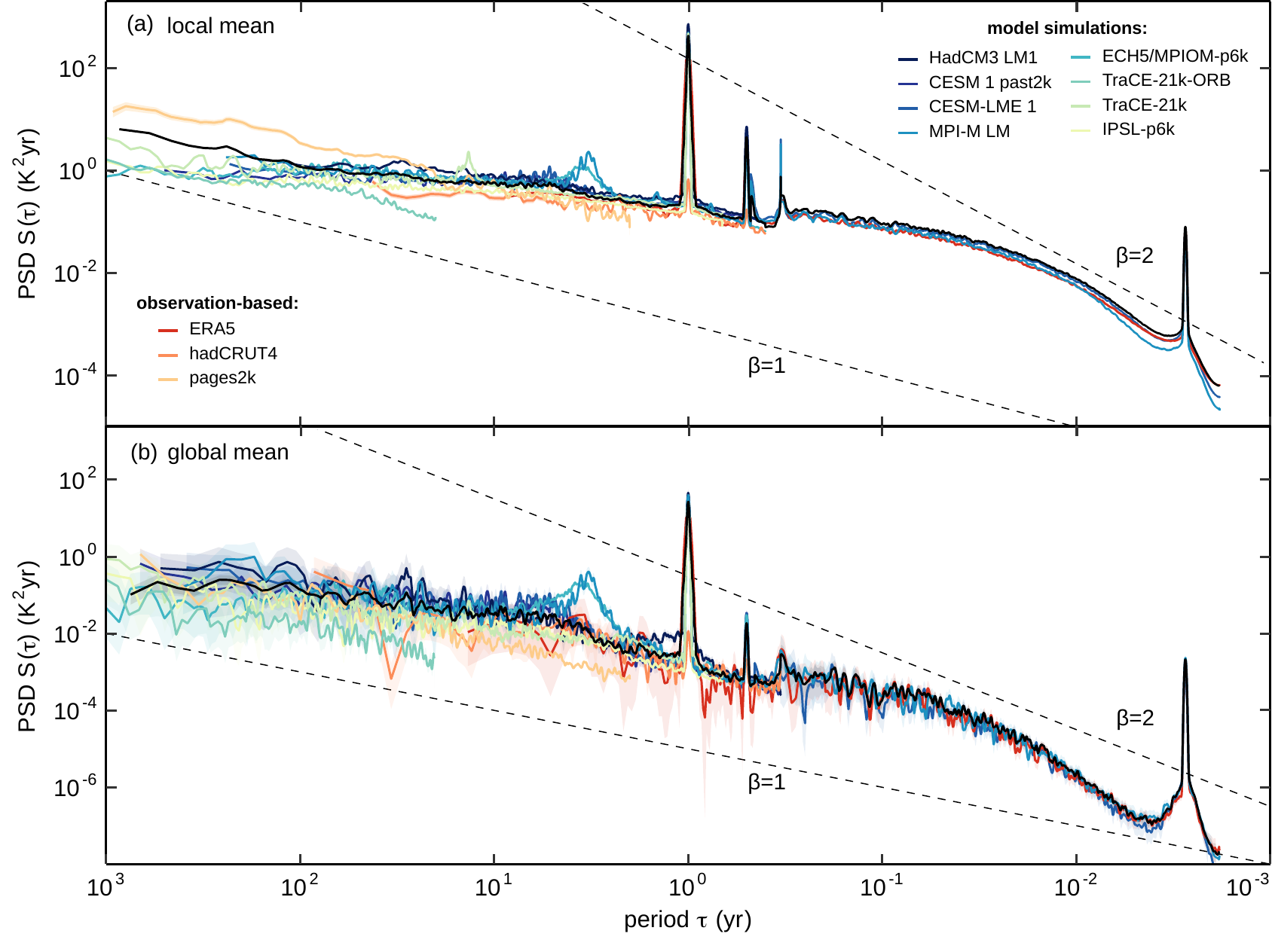}%
	\caption{\label{fig:gmstrmst} (a) Mean power spectral densities (PSD) of local temperature from model simulations and observation-based data on periods from hours to 1000 yr for the Holocene. (b) PSD of global mean temperature. The dashed lines with slope $\beta$ and arbitrary $y$-intercept in the log-log graph indicate the scaling behavior for visual comparison. The ensemble means (black solid lines) were formed using equal weights across the model group M$_{0}$ (see Table S1 \cite{Supplementary}).} 
\end{figure*}

In order to study the timescale dependency of global mean temperature, we present its power spectral density in Fig.~\ref{fig:gmstrmst} (b). It shows the characteristic background continuum, spectral peaks, and higher harmonics associated with the diurnal and annual cycle. Overall, the PSDs tend to agree between the data sets, albeit with some differences on the interannual scale and when compared to the Trace21k ORB run. The Trace21k-ORB run is solely forced by orbital changes and therefore shows less variability than the ensemble mean. The broad spectral peak on interannual periods reveals an artificially amplified ENSO in the shared MPI-M LM and ECHAM5/MPI-OM ocean component \cite{Jungclaus2020}. For a better visibility, PI control runs are separately shown in the supplementary Fig. S6 \cite{Supplementary}. Overall, the PSD largely agrees among different data sets, especially towards shorter timescales.

We find a power-law scaling of $\beta \approx 1$ on timescales longer than 10 yr in line with previous results \cite{Lovejoy2015, Rypdal2013, Nilsen2016}. The PSD decreases more strongly towards shorter periods, which is characteristic of the weather regime \cite{Lovejoy2015, Pelletier2002}. Similar to Nilsen \textit{et al.} \cite{Nilsen2016}, we find no evidence for significant changes in scaling behavior around the centennial scale. One limitation of previous work that found scale breaks is that the spectra were estimated across nonstationary shifts in climate, such as the deglaciation \cite{Zhu2019a}, and with a change in proxies and archives \cite{Huybers2006}.

We present the area-weighted mean spectra of the local (grid box) temperature in Fig.~\ref{fig:gmstrmst}(a). Compared to the global mean in Fig.~\ref{fig:gmstrmst}(b), the power increases and the spectral slope decreases, in line with \cite{Fredriksen2016}. The spectra agree on periods below 10 yr, except for the artificially amplified ENSO signal mentioned earlier. Moreover, we find a narrow peak at 13 yr, associated with an unrealistic variability in the northern North Atlantic of the TraCE-21k run, similar to \cite{Danabasoglu2008, Kunz2021}. Remarkably, the decadal-to-centennial variability of the reconstructed temperature is increased by one to two orders of magnitude compared to the simulations. The spectral exponent is smaller for models ($\beta < 1 $) compared to paleoclimate data  ($\beta \approx 1$). 

This finding verifies that models show less regional temperature variability and that the mismatch increases towards longer timescales. The results are robust to sampling from the PAGES2k database and the influence of anthropogenic climate change (Fig. S10 \cite{Supplementary}). One shortcoming of forming the area-weighted mean PSD is that the uncertainty quantification requires the assumption of independent spatial degrees of freedom of the temperature field. Due to the presence of spatial correlations, an estimate of the effective spatial degrees of freedom and their dependence on the underlying timescale would be needed to resolve this limitation \cite{Kunz2018}. 

\subsection{\label{sec:scalingmap2}Spatial patterns of persistence}

\begin{figure*}[!htbp]
\includegraphics[width=13.2cm]{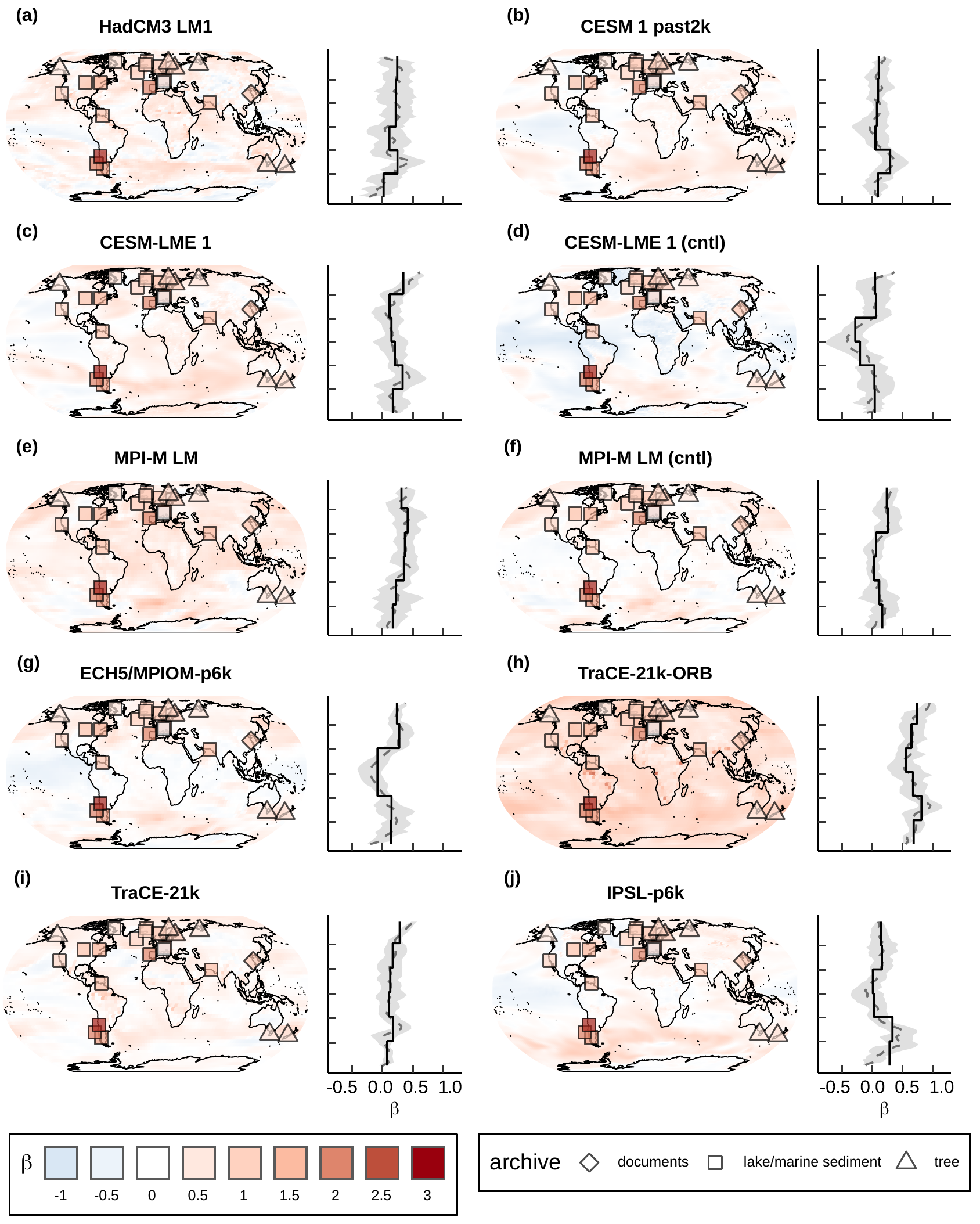}
\caption{\label{fig:scalingmap} 
Local temperature persistence on timescales from $\tau_1= 10$ to $\tau_2=200$ yr across multiple climate simulations and selected proxy records from the PAGES2k database. Colors from blue to red indicate the scaling behavior ranging from $\beta=-1$ to $\beta=3$. Symbols indicate the scaling of proxy records from different natural archives. The background of each panel shows the $\beta$-values fitted to the PSD of the local grid box temperature from simulations. Zonal mean values (dashed curves) are given next to the map, with means (solid curves) over latitude-intervals (with breaks at $-60$, $-30$, $0$, $30$, and $60^\circ$ N) and gray shaded confidence intervals. The spatial coverage of proxy records is not sufficient for robust mean estimates, which is why only simulation data are shown here. }
\end{figure*}

To further investigate the mismatch on local scaling properties, we compare the spatial dependence of temperature persistence from simulations and paleoclimate data in Fig.~\ref{fig:scalingmap}. The simulations largely exhibit small-magnitude scaling exponents ($-1< \beta < 1$), whereas proxy records were found to also show $\beta > 1$. In this manner, the magnitude of local temperature fluctuations from model simulations often shows no dependence on the decadal-to-centennial timescale. However, approximately half of the proxy records show a variance that grows on increasingly long periods (see also Fig. S11 \cite{Supplementary}). 

From both simulations and paleoclimate data, we can strengthen the argument by Fredriksen \textit{et al.} \cite{Fredriksen2016} that there is no latitudinal dependence of $\beta$ (Fig.~\ref{fig:scalingmap}), in contrast to previous studies, suggesting a possible linkage to the strength of the seasonal cycle \cite{Huybers2006}. Inspecting the simulations' $\beta$-values (background of Fig.~\ref{fig:scalingmap}), we find a small land-sea contrast. Strongest scaling occurs in the Southern Oceans in line with previous findings \cite{Fredriksen2016}. Ocean-sea ice interactions with characteristic timescales of the order of centuries and a generally increased internal variability over the oceans might explain these results. 

We find generally lower values for the slope $\beta$ in the ENSO and Indo-Pacific region. This could be attributed to the fact that (quasi-)oscillatory signals, such as active modes of internal variability, are reflected in the PSD as broad peaks and hence cannot be described by a scaling law. On the other hand, this finding is stronger in PI control runs compared to fully forced runs [Fig.~\ref{fig:scalingmap} (c-f)]. Thus, residual effects of the recent global warming trend might play an additional role \cite{Yeh2009}. A systematic bias becomes clear from the spatially almost uniform $\beta$-values of Trace21k-ORB [Fig.~\ref{fig:scalingmap} (h)]. In line with Fig.~\ref{fig:gmstrmst}, we explain this by the lack of forcing mechanisms on interannual to multi-decadal timescales in the aforementioned simulation. 

Marine and lake sediments, as well as the archived documents, follow the general trend of increased $\beta$-values compared to simulations. Tree ring records agree well with most simulations in North America and Siberia, but not necessarily at the coast of Australia and northern Europe. Discrepancies such as those in southern South America could reflect the proxies' strength in representing local conditions, for example, topography. However, noise sources in the climate signal recording and preservation, such as bioturbation, can influence proxy records. Further separating the signal content from noise sources in paleoclimate reconstructions can help refine our findings \cite{Reschke2019, Lucke2019}.

\subsection{\label{sec:statistics}Statistical agreement of temperature persistence}

\begin{figure}[!htbp]
	\includegraphics[width=8.6cm]{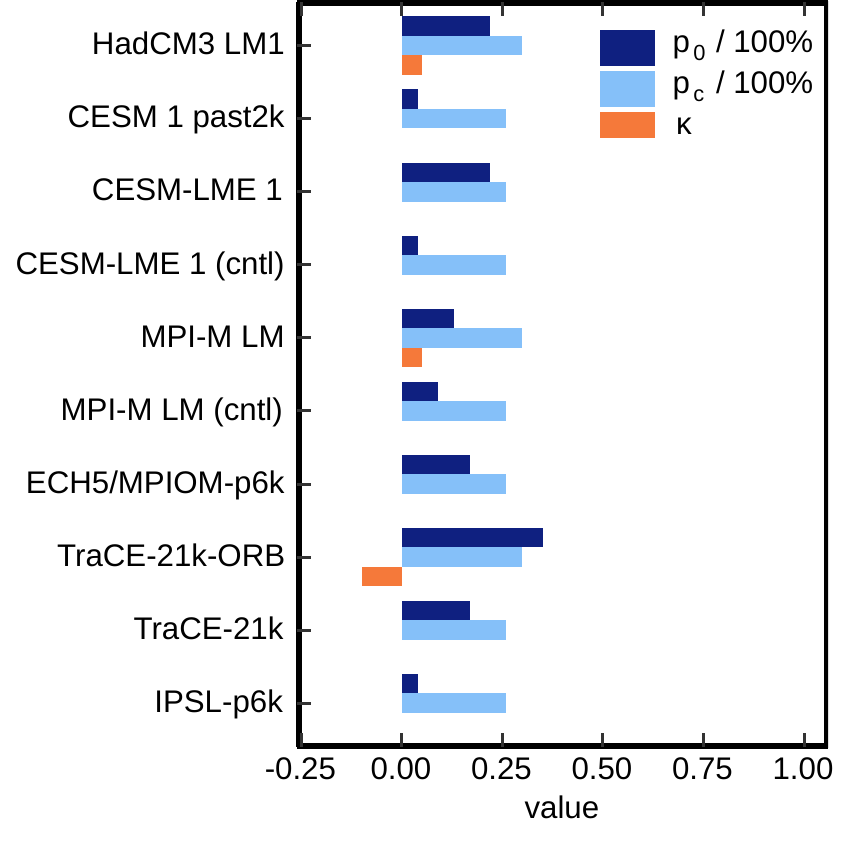}
	\caption{\label{fig:statistics} Percentage agreement $p_0$, categorical agreement $p_c$ and inter-rater reliability $\kappa$ of local temperature persistence from simulations and paleoclimate data. The measures were calculated from a set of bilinearly interpolated simulation records and the proxy record at 23 different locations. Missing orange bars indicate no agreement beyond chance and, therefore, zero inter-rater reliability ($\kappa=0$).}
\end{figure}

We further investigate the question of temperature scaling by a statistical analysis of $\beta$-values from simulations and paleoclimate reconstructions. It is based on the detailed uncertainty quantification outlined in \autoref{sub:uncertainties}. Our results show that reconstructions and simulations agree in less than 30\% of locations within the scope of uncertainties (Fig.~\ref{fig:statistics}). To single out the scaling behavior of temperature signals, we study the agreement by category. We find approximately 25\% of agreement within the categories $\beta < 1 - \nu$ (\textit{low}) and $\beta > 1 + \nu$ (\textit{high}). Although widely accepted \cite{Fleiss2013}, categorical and percentage agreement suffer from the limitation to ignore any agreement by chance. Therefore, we investigate the $\kappa$-statistics (orange bar in Fig.~\ref{fig:statistics}) and verify that there is no agreement beyond chance ($\kappa=0$) for almost all models. Only MPI-M LM and HadCM3 LM1 show any, if poor agreement ($\kappa \approx 0.1$), whereas Trace21k-ORB shows even lower agreement than expected by chance ($\kappa < 0$) due to its systematic bias. 

The disagreement could be attributed to both paleoclimate data and simulations. A systematic bias could arise, for example, through the recent, nonstationary global warming trend. Therefore, we repeat our analysis with all time series cut at 1850. In particular, anthropogenic warming slightly increases long-term temperature variability and thus scaling behavior, but not significantly (Figures S6, S9, and S10 \cite{Supplementary}). Further uncertainties could arise from our choice of statistical estimator for the scaling exponent $\beta$. Maximum likelihood estimation (MLE) should generally be preferred over linear regression (LR) because of its mathematical soundness and skillfulness \cite{Clauset2009}. We find that MLE is indeed more accurate for regular time series with $\beta>0$ (Fig. S14 \cite{Supplementary}). However, LR allows for estimation of $\beta <1$, unlike MLE which assumes $\beta >1$ \cite{Clauset2009}. In addition, for the characteristics of our empirical data, the differences between the two methods are not significant for $\beta >0$ (Fig. S15 \cite{Supplementary}). Therefore, linear regression represents the preferred estimator for our analysis. Regardless of the chosen estimator, we observe a slight tendency towards increased scaling exponents for irregularly sampled data (Fig. S15 \cite{Supplementary}), similar to Lucke \textit{et al.} \cite{Lucke2019}. Our uncertainty quantification carefully accounts for these potential errors due to irregular sampling and interpolation by simulating their influence using surrogates (Fig. S8 \cite{Supplementary}).

We do not expect other systematic biases for the paleoclimate data since we base our results on multiple archives and proxies, and no systematic spatial pattern is discernible (Fig. S11 \cite{Supplementary}). In particular, the cross-correlations between the 23 proxy datasets are weakly positive (0.02 on average with 95\% quantiles of -0.17 to 0.21). The assumption of spatial independence necessary for robust statistical analysis (Fig. S16 \cite{Supplementary}) therefore appears fully satisfied. The models' resolutions are another possible element of uncertainty that impacts variability over a wide range of timescales \cite{Kirtman2012, Klavans2017, Hodson2012}. We here facilitate inter-model comparison by using state-of-the-art GCMs with comparable spatial and temporal resolutions, but computational costs precluded higher resolutions. The latter might be necessary to improve the representation of decadal variability and response to external forcing. In particular, the increased scaling exponents ($\beta > 1$) from paleoclimate data could indicate that nonlinear processes from an interactive carbon cycle and dynamical ice sheets might not be sufficiently represented in models. 

\subsection{The forced temperature response}

\begin{figure}[!htbp]
	\includegraphics[width=8.6cm]{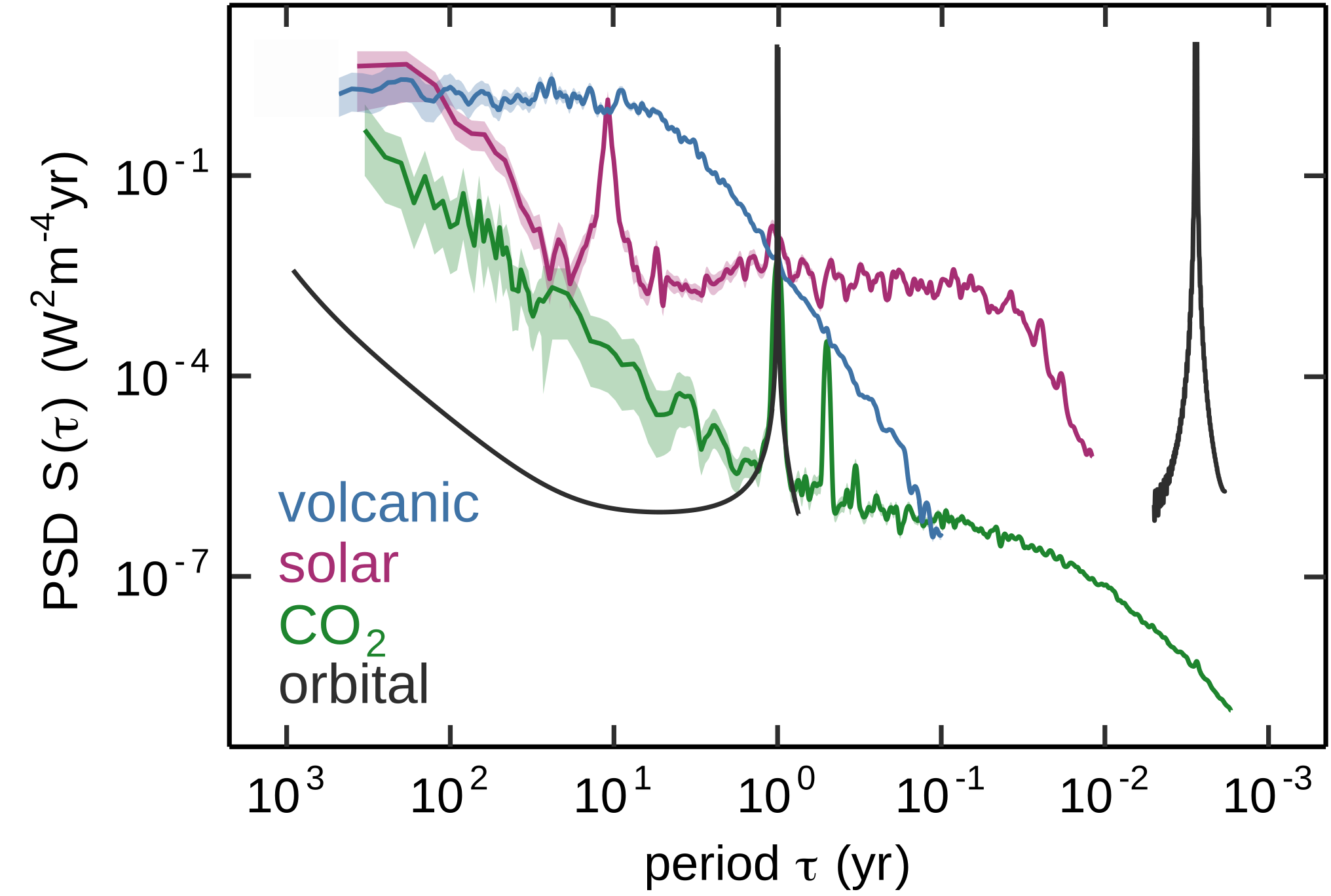}%
	\caption{\label{fig:forcings_specs_mean}
		Power spectral densities from radiative forcings. Details on the reconstructions considered here are summarized in Table S2 and Figure S5 \cite{Supplementary}.
	}
\end{figure}

Climatic drivers are not constant in time and thus affect the surface air temperature on multiple timescales. To investigate the forced temperature response, we present spectra for the main climatic drivers in Fig.~\ref{fig:forcings_specs_mean}. The PSD of orbital forcing consists of the diurnal and annual cycle as well as a background continuum on longer timescales. Higher harmonics on monthly timescales were omitted. We calculate the mean volcanic, solar, and CO$_2$ spectra using an equally weighted average of spectra from multiple data sets (Fig. S5 \cite{Supplementary}). The CO$_2$  forcing follows the orbital forcing. The PSD of solar forcing again contains more power and has a pronounced peak around the 11 yr solar cycle. Multiple theories and paleoclimate reconstructions suggest the increased variability on centennial to millennial periods due to the long-term behavior of solar activity \cite{Gray2010}. 

\begin{figure}[!htbp]
\includegraphics[width=8.6cm]{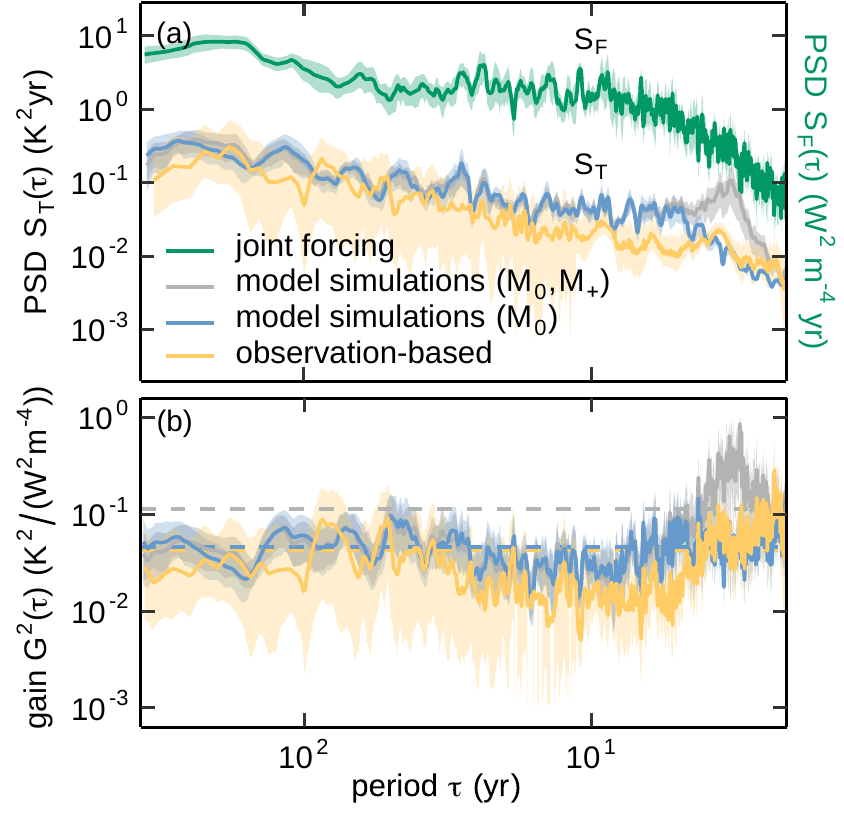}%
\caption{\label{fig:gain} Monte Carlo simulation of
PSD (a) and spectral gain (b)  using temperature and forcing reconstructions as well as model simulations. Shaded confidence intervals lie between the 5\% and 95 \% quantiles. We consider only models from the groups M$_{0}$ and M$_{+}$ (Table S1 \cite{Supplementary}) to exclude model artifacts and to represent the historical temperature response in the best possible way. Notably, M$_{+}$ contains those simulations with amplified ENSO \cite{Jungclaus2020}. (b) Dashed lines indicate the mean variance ratio $\langle{S_T}\rangle/\langle{S_F}\rangle$.}
\end{figure}

Volcanic forcing dominates interannual to centennial scales and undergoes a scale break around the period of 7 yr, estimated using the goodness of fit \cite{Clauset2009}. Above decadal scales, it follows a white noise spectrum with constant variance. However, the intermittency of volcanic eruptions might have led to biases in the spectral characteristics \cite{Lovejoy2016}. We verify our results using an analytical approach described in Appendix \ref{app:forcingestimation}. Remarkably, the derived PSD of an ideal, intermittent time series with Poisson distributed return times explains our findings. We further demonstrate the scale break by a Monte Carlo simulation of the joint PSD of radiative forcing in Fig.~\ref{fig:gain} (a). This finding raises the question of how the spectrum with a scale break translates into the continuous spectrum in Fig.~\ref{fig:gmstrmst} (b).

We address this question by calculating the spectral gain \eqref{eq:gain} on periods between years and centuries in Fig.~\ref{fig:gain} (b). Here, observation-based data include HadCRUT4, ERA5, and PAGES2k again. To account for the model artifacts explained above, we calculate the gain from the model simulation group M$_{0}$ and together with group M$_{+}$ (see Table S1 \cite{Supplementary}). We find that the spectral gain is similar from observation-based data and the model simulation group M$_{0}$, which is the one without artificially amplified ENSO. This suggests that both follow a similar distribution of timescale-dependent variability, as already indicated by Fig.~\ref{fig:gmstrmst}(b). Large parts of the gain show constant behavior, which is most pronounced in M$_{0}$. In a simplified way, the gain might be approximated by an ideal linear amplifier or damper of the forcing with comparable internal variability on all timescales. However, we also find a dip around decadal scales, which is strongest in the gain from measurements. Inspecting Fig.~\ref{fig:gain} (a), this can be explained by forming the ratio between a spectrum with a scale break ($\beta > 1 \rightarrow \beta \approx 0$) and one with moderate scaling ($\beta \approx 1$). 

From this standpoint, internal variability slightly grows on periods from years to centuries when slow processes in the oceans, vegetation, land surface, and cryosphere become increasingly active (Fig.~\ref{fig:climatesystem}). While the model simulations follow this general pattern, they may not represent its amplitude correctly, for example, due to the lack of feedback mechanisms. In addition, a too high model diffusivity could cause the suppression of low-frequency variability in model simulations due to a faster energy dissipation over temporal scales \cite{Laepple2014GRL}. The PAGES2k multiproxy reconstruction, stemming from palaeoclimate data, possibly underestimates internal variability on interannual scales. However, the mean variance ratios in Fig.~\ref{fig:forcings_specs_mean} (b) of the model estimates agree with those from observations in the global mean. This leaves us with a conundrum: the global mean temperature based on model simulations and observations is mostly consistent in its variability, scaling, and response to forcing. Notwithstanding, locally, the models show a much lower variance on longer timescales and different scaling behavior than reconstructions. Thus, it appears that the statistics of local fluctuations need to be optimized in models but without significantly altering global properties. To this end, the study of unforced (``spontaneous'') oscillations \cite{Hertwig2015} and abrupt transitions \cite{Peltier2014, Klockmann2020} in the climate system is one promising approach to improve the representation of local variation. Furthermore, higher-resolved ocean and atmosphere models with additional mechanisms such as ice sheet dynamics and an interactive carbon cycle might increase long-range dependence and persistence of local temperature in the future.

\section{\label{sec:conclusion}Conclusion} 

In summary, we have investigated the question of temperature variability on the timescale of years to centuries. To this end, we have presented power spectral densities for both local and global surface air temperature from simulation and observation-based data of the last millennia. On this basis, we concluded that locally there is a stronger scaling and increased variance in reconstructions as compared to simulations. Using statistical analysis, we found that local temperature series extracted from simulations and paleoclimate reconstructions show different scaling behavior, with proxy records hinting at a stronger persistence. Furthermore, we have largely extended the spectral analysis of climatic drivers by estimating the joint PSD from CO$_2$, solar, volcanic, and orbital forcing using Monte Carlo simulation. Hereby, we discovered a scale break at the period of approximately 7 yr. Moreover, we have presented the spectral gain, describing the timescale-dependent forced temperature response. We found that it is mostly consistent across data sets and indicates an increasing internal variability on timescales of decades to centuries.

Our analysis of the spectral gain was limited to global average values and those timescales where linearity can be reasonably assumed \cite{RypdalRypdal2016, Lovejoy2016, Fernandez-Donado2013}. Nonlinearities are inherent to the climate system, for example, due to the temperature-albedo feedback. Thus, it will be necessary to examine their possible effects on multiple spatiotemporal scales to further extend this work. Studying nonlinearities could also shine new light on the mechanisms of scaling in Earth's climate, which are not yet fully understood and might be linked to nonlinearities as well \cite{Franzke2020}. Furthermore, we have focused on the current interglacial, the Holocene. This is because climate variability has been demonstrated to depend on the mean climate state \cite{Rehfeld2018}. Furthermore, major shifts in climate could potentially violate the basic assumption of weak stationarity for spectral analysis. Thus, the conclusions laid out here cannot be readily applied to other climate states, such as glacial periods, which is an issue for future studies. Clearly, understanding the dependence of temperature variability on global warming demands additional work.

Ideally, our findings should be replicated by employing models with increased internal variability on longer timescales and paleoclimate data that provides improved spatiotemporal resolution. In particular, investigating the relationship between spatial and temporal disagreement is a key task for future analyses. Optimized analysis of noise sources and spectral analysis of (pseudo-)proxy records could help to expand the data basis of proxy records with decadal resolution \cite{Kunz2020, Dolman2020, Casado2020}. Regarding climate models, an improved representation of processes that increase Earth's long-term memory, such as an interactive carbon cycle and dynamical ice sheets, might strengthen the long-range dependence and persistence of surface air temperature. A better understanding of unforced low-frequency oscillations as well as abrupt changes will be necessary to improve the representation of local fluctuations and could further help to understand nonlinear feedback and possible bifurcations in the climate system. Future studies could also continue to explore how internally generated and externally  forced variability compares on different spatial scales. Research on the interrelation between internal and forced changes, as well as local, regional, and global variability, might prove important and could be conducted using single-forcing experiments from ensembles of model simulations.

Managing climate risks requires a detailed understanding of temperature variability. Locally and on timescales between years and centuries, there is an urgency to address discrepancies to make further progress in climate modeling. In this study, we have singled out the key characteristics of temperature variability and showed that the timescale dependency of local temperature variations from observation-based data and model simulations differs. Our results have demonstrated that the scaling behavior and spectral gain are easy-to-use yet effective and promising tools for investigating variability in Earth's dynamic climate.

Code to reproduce all figures is available at \cite{code}.

\begin{acknowledgments}
This manuscript is based upon data provided by the World Climate Research Programme’s Working Group on Coupled Modelling, which is responsible for CMIP and PMIP. We thank the research groups listed in Tables S1 and S2 for producing and making available their data from model outputs, measurements, paleoclimate, and forcing reconstructions. This study benefited from discussions within the CVAS working group, a working group of the Past Global Changes (PAGES) project. We thank T. Gasenzer, T. Kunz, and N. Weitzel for discussions and J. Bühler, M. Casado, M. Schillinger, and E. Ziegler for helpful comments on the manuscript. We are grateful to Aimé Fournier and one anonymous referee for their constructive and valuable review. This research has been funded by the Heidelberg Graduate School for Physics, by the PalMod project (subproject no. 01LP1926C), and by the Deutsche Forschungsgemeinschaft (DFG, German Research Foundation) – project no. 395588486. 
\end{acknowledgments}

\appendix

\section{\label{app:PowerSpectrum} Relation between power spectral density and variance}

The power spectral density of a weakly-stationary, stochastic process is given by the Fourier transform of the autocorrelation $S(f)= \mathcal{F}\lbrace R(h)\rbrace$ with frequency $f$ and lag $h=t_2 - t_1$ between two points in time \cite{Wiener1930, Khintchine1934}. For zero lag and zero mean, the integral of the PSD corresponds to  the variance of the signal \cite{Chatfield2019}. Instead of frequency, we use the period $\tau=1/f$ to express the PSD and spectral gain. The integration of expression (1) is divergent for $\beta <1$ and $f \rightarrow \infty$ which requires a high-frequency cut-off, such as described by Lovejoy \textit{et al.} \cite{Lovejoy2019}. In case of temperature time series considered here, this is naturally defined by the temporal resolution, setting the maximum frequency.

\section{\label{app:Autocovariance} Autocovariance of long-range memory processes}

Fractional Brownian motion (fBm) and fractional Gaussian noise (fGn) are fully described by their correlation properties \cite{Mandelbrot1968, Fournier2021}, summarized below. 
The autocovariance function of fBm $B(t)$ reads
\begin{align}
    \label{eq:covariance_fbm}
    \gamma (t', t) = \langle B(t')B(t) \rangle &= \frac{V_\beta}{2} \left( \vert t \vert^{\beta - 1} + \vert t' \vert^{\beta - 1} - \vert t' - t \vert^{\beta - 1}\right) \nonumber \\
    &\propto 1 + \Big\vert \frac{t'}{t}  \Big\vert^{\beta - 1} - \Big\vert 1- \frac{t'}{t}  \Big\vert^{\beta - 1} 
\end{align}
for $1 < \beta < 3$. 
$V_\beta$ is a positive constant factor related to $\langle (B(t')-B(t))^2 \rangle = V_\beta \, \vert t' - t \vert^{\beta - 1}$. 
By definition, fGn is the series of stationary increments $B(t') - B(t)$ and shows spectral exponent $-1 < \beta' = \beta-2 < 1$ for $f \ll 1/\pi \Delta t$ with $\Delta t=t'-t$. Its autocovariance 
     \begin{align}
        \label{eq:covariance_fgn}
       \gamma(h) &= \langle (B(t+ 1 +h) - B(t+h))(B(t+ 1)-B(t))\rangle \nonumber \\
       &=   \frac{V_\beta}{2} \vert h - 1 \vert^{\beta' + 1} - 2 \vert h \vert^{\beta' + 1} +  \vert h + 1 \vert^{\beta' + 1} \,,
    \end{align}
depends only on the lag $h\in\mathbb{Z}$, where we set $\Delta t= 1$ without loss of generality. The fGn has a power spectrum of the form \cite{Mallat2009}
    \begin{equation}
    \label{eq:fGNspectrum}
        S(f) \propto \frac{\sin^2(\pi \Delta t f )}{ \vert 2 \pi \Delta t f \vert^{\beta'+2}} \,, 
    \end{equation} 
    with the slowly varying factor
    \begin{equation}
        \sin^2 (\pi \Delta t f) \underset{f/f_\mathrm{max} \rightarrow \, 0}{\xrightarrow{\hspace*{1.2cm}}} (\pi \Delta t )^2 f^2, \qquad f_{\mathrm{max}}=1/\pi\Delta t \,.
    \end{equation}
Considering positive frequencies $f>0$, the spectrum \eqref{eq:fGNspectrum} can be approximated by the power law $S(f) \sim 1/ f^{\beta'}$ if $f \ll f_{\mathrm{max}}$. For $f \gtrsim f_{\mathrm{max}}$, however, the fGn has a similar spectral shape to fBm \cite{Fournier2021}. We account for this by considering sufficiently long periods. To give an example, $10^{0.58}$ yr$^{-1} \lesssim f_{\mathrm{max}} \lesssim 10^{2.7}$ yr$^{-1}$ corresponds to $6$ h $\lesssim \, \Delta t \, \lesssim 1$ mon.

For all $\vert t' / t \vert \gg 1$, the covariances \eqref{eq:covariance_fbm} keep growing for $\beta>2$ (persistence) and stay bounded for $\beta < 2$ (antipersistence). 
As a result, equation \eqref{eq:covariance_fbm} involves ``nonlinear pseudo-trends'' \cite{Mandelbrot1968} for $B(t')$ conditioned on $B(t)$, which diverge for $\beta > 2$ and converge for $\beta < 2$. 
According to Eq. \eqref{eq:covariance_fgn}, fGn is persistent for $\beta' > 0$ and antipersistent for $\beta' < 0$. 
Ordinary Brownian motion corresponds to $\beta=2$ and white noise to $\beta'=0$. 
The sequence of partial sums of the autocovariance function diverges for fGn with $\beta'>0$ and fBm with $\beta>2$. The process is nonsummable and said to possess long-range memory.

\section{\label{app:Gain} Spectral gain for linear systems}

In a time-invariant linear system, the output
\begin{equation}
    \label{eq:response}
    y(t) = \int_{-\infty}^{\infty} h(u) x(t-u) \mathrm{d}u 
\end{equation} 
is given by the input time series $x(t)$ and the impulse response function $h(u)$ \cite{Chatfield2019}. The Fourier transform $H(f)= \mathcal{F}\lbrace h(u) \rbrace=G(f) e^{i \phi(f)}$ gives the frequency response function, also called the transfer function. $G(f)$ and $\phi(f)$ are the gain and phase, respectively. The integral \eqref{eq:response} corresponds to a product in frequency space $\mathcal{F} \lbrace y(t) \rbrace  = H(f) \mathcal{F} \lbrace x(t) \rbrace$. This relates the PSD of the output $S_y(f)$ to the one of the input $S_x(f)$ via
\begin{equation}
S_{y}(f) = \vert H(f) \vert ^2  S_{x} (f)  = G^2(f) S_{x} (f)  \,.   
\end{equation}

\section{\label{app:forcingestimation} Analytical solution to the PSD of intermittent volcanic forcing}

We investigate the power spectral density of intermittent volcanic forcing by approximating the eruption time series in a simplified way as a stochastic signal $X(t)=\delta(t - t_i)$. This function is zero at all times except $t_i$, when an event of unique amplitude occurs. We denote $T_i= t_{i}-t_{i-1}$ the time intervals between two events. We use the fact that the PSD cannot be calculated only from the covariance, but also from the Laplace transform $S(X, f)=2 \lim_{\epsilon \to 0} \langle \vert \mathcal{L}(X(t), \frac{\epsilon}{2} -  2 \pi i f) \vert^2 \rangle $\cite{Stratonovich1967}. Based on this approach, the power spectral density
\begin{equation}
\label{eq:SofPorbdensity}
     S(f)= \mu_T\dfrac{1-\vert \rho (f) \vert^2 }{\vert 1- \rho (f) \vert^2} \,,\, f > 0
\end{equation}
becomes a function of the Fourier transform of the probability density function $\rho(f)= \mathcal{F}\lbrace\rho (T)\rbrace$ and the inverse mean interval between two events $\mu_T=\langle T \rangle^{-1}$ \cite{Stratonovich1967, Lindner2006}. An exponentially decaying probability distribution $\rho(T)= \mu_{T}\,\mathrm{exp}(-\mu_T T)\Theta(T)$ for volcanic forcing is suggested \cite{Papale2018}, and we have checked this for the data sets considered. The Fourier transform reads $\rho(f)=\mu_T (\mu_T+2\pi i f)^{-1}$ such that $1-\vert \rho(f) \vert^2 = \vert 1- \rho(f) \vert ^2$. As a consequence, the PSD \eqref{eq:SofPorbdensity} takes a constant value. We can observe this white noise behavior in Fig.~\ref{fig:forcings_specs_mean} and \ref{fig:gain} (a) on timescales longer than a few years, which is on the order of characteristic return times for eruptions. Below these timescales, the variability considerably drops. 
This analytical result provides an independent verification of the PSD for volcanic forcing and its scale break. 

\section{\label{app:MonteCarlo} Monte Carlo sampling of the spectral gain}

We simulate the spectral gain \eqref{eq:gain}, as well as the PSD of global mean temperature and the joint PSD of radiative forcing using a Monte Carlo approach with $N=1000$ realizations to account for sampling biases.
The PSD of global mean temperature is sampled for three groups: the observation-based data, the model simulations from group M$_{0}$, and those from M$_{0}$ together with M$_{+}$ (Table S1 \cite{Supplementary}). Here, only models from the groups M$_{0}$ and M$_{+}$ are considered to exclude model artifacts and to represent the historical temperature response in the best possible way. 

We sample the simulation-based PSD from the average PSD of the simulations using uniformly distributed random weights. 
To obtain the observation-based PSD, we use the global mean temperature from HadCRUT4, ERA5, and a 7000-member reconstruction ensemble provided by PAGES2k \cite{Neukom2019}. This ensemble allows us to sample the PSD by randomly selecting one ensemble member and form the mean of its spectrum with that of the ERA5 and HadCRUT4 temperature. 
The joint PSD of radiative forcing is calculated from all forcing reconstructions considered in this work except the Fröhlich \textit{et al.} solar forcing, which has too low temporal resolution above interannual scales (Table S3 and Fig. S5 \cite{Supplementary}). We assume the PSD of CO$_2$ and orbital forcing as fixed since its spectral power is comparatively low on multi-decadal scales. We sample the PSD of solar forcing by using uniformly distributed weights when forming the average PSD of all solar reconstructions. Similarly, the PSD of volcanic forcing is obtained. In addition, we randomly vary the conversion factor between $(-18)^{-1}$ and $(-25)^{-1}$  $\mathrm{W m^{-2}}$/AOD \cite{IPCC2013}. The joint PSD of radiative forcing is calculated by linear summation of the PSD from CO$_2$, orbital, solar, and volcanic forcing.

Using this sampling scheme, our Monte Carlo produces two outcomes: First, we compute the PSD of global mean temperature and the joint PSD of radiative forcing by simulating an ensemble of $N$ realizations for both forcing and response. Second, we sample the spectral gain directly from the quotient \eqref{eq:gain} in each of the $N$ realizations. In both cases, the average of the generated $N$-member ensemble and its 5\% and 95\% quantiles constitute the result of our Monte Carlo simulation.

\end{document}


\title{Supplemental Material for \\ 
\textit{Probing the timescale dependency of local and global variations in surface air temperature from climate simulations and reconstructions of the last millennia}}
\author{Beatrice Ellerhoff and Kira Rehfeld}

\maketitle
\nopagebreak

\listoftables
\listoffigures

\pagebreak
\begingroup
\squeezetable
\begin{table*}[!htbp]
\caption[\,Key specifications of model simulation and observation-based data]{\label{tab:signal_data} Key specifications of model simulations and observation-based data used to estimate temperature variability. We give the main references (Doc.) of the simulation runs, observation-based data sets, and their forcing (Forc.). The groups M$_{+}$ and M$_{0}$ were assigned to better distinguish spectral properties when performing a Monte Carlo simulation of the spectral gain in the main manuscript. The temporal resolution $\Delta t$ is given in months (m), hours (hr), years (yrs), and days. The spatial resolutions of atmosphere and ocean are denoted by the subscripts $^{\circ}$deg$_{at.}$ and $^{\circ}$deg$_{oc.}$, respectively.
}
\begin{ruledtabular}
\begin{tabular}{llllllll}
Name & Doc. & Model & $\Delta t$ & $^{\circ}$deg$_{at.}$ & $^{\circ}$deg$_{oc.}$ & Forc. & Time (CE) \\ \hline
\multicolumn{8}{l}{\textbf{Model simulations}} \\
CESM 1 past2k\footnote{assigned to group M$_0$} & \cite{Zhong2018}  & CESM1 & 1 m & 2 & 1 &  \cite{Jungclaus2017} & 1-2005  \\ 
CESM-LME 1\footnotemark[1]& \cite{Otto-Bliesner2016}  & CESM1 & 1 m, 6 hr & 2 & 1 &  \cite{Schmidt2011} & 850-2006  \\ 
CESM-LME 1 cntl & \cite{Otto-Bliesner2016}  & CESM1 & 1 m & 2 & 1 &  \cite{Schmidt2011} & 850-2006   \\ 
MPI-M LM\footnote{assigned to M$_{\mathrm{+}}$} & \cite{Jungclaus2010} & MPI-ESM & 1 m, 6 hr & 3.75 & GR30\footnote{curvilinear grid with nominal resolution of 3.0$^{\circ}$} &  \cite{Jungclaus2010} & 800-2005  \\ 
MPI-M LM cntl & \cite{Jungclaus2010} & MPI-ESM & 1 m & 3.75 & GR30\footnotemark[3] &  \cite{Jungclaus2010} & 800-2005  \\ 
HadCM3 LM1\footnotemark[1] & \cite{Buehler2020} & iHadCM3 & 1 m & 2.5 x 3.75 & 1.25 &\cite{Schurer2014} & 800-1850   \\ 
IPSL-p6k\footnotemark[1] & \cite{Braconnot2019} & IPSL-CM5A  & 2 m & 2.5 x 1.27 & 2 & \cite{Braconnot2019} & 4000 BCE - 2000  \\
TraCE-21k & \cite{Liu2009} & CCSM3 & 2 m & $\approx 3.75$ & 3.6 x v\footnote{the latitudinal resolution is variable (v), with finer resolution near the equator ($ \approx 0.90 ^\circ$)}  &   \cite{Berger1978, Joos2008} & 4,000 BCE - 1990 \\
TraCE-21k-ORB  & \cite{Liu2009} & CCSM3 & 10 yrs  & $\approx 3.75$ &  3.6 x v\footnotemark[4] &  \cite{Berger1978, Joos2008} & 4,000 BCE - 1990  \\
ECH5/MPIOM-p6k\footnotemark[2] &  \cite{Fischer2011} & ECHAM5/MPI-OM & 1 m & 3.75 & 2 & \cite{Berger1978}  & 4000 BCE - 2000 \\ \hline
\multicolumn{8}{l}{\textbf{Observation-based}}  \\ 
ERA5 & \cite{Hersbach2020} & & 1 m, 6 hr & 2 & 2 & &  1979-2019 \\ 
HadCRUT4 & \cite{Morice2012} & & 7 days & 5 & 5 &  &  1850-2019 \\ 
PAGES2k & \cite{Neukom2019} & & 1 yr &  &  & \cite{Neukom2019} &  0-2000 \\ 
\end{tabular}
\end{ruledtabular} 
\end{table*}
\endgroup
\begin{table}[!htbp]
\caption[\,Key specification of climatic drivers]{\label{tab:forcing_data} Key specifications of climatic drivers used to estimate the power spectral density of radiative forcing as well as the gain function of the forced temperature response.}
\begin{ruledtabular}
\begin{tabular}{lll}
Name & $\Delta t$  & Time (CE)            \\ \hline
\multicolumn{3}{l}{\textbf{Volcanic forcing}} \\
Crowley et al.\footnote{from the PMIP simulations of the Last Millennium \cite{Schmidt2012}} \cite{Crowley2013} & 10 days & 500BCE-1900 \\ 
Gao et al.\footnotemark[1] \cite{Gao2008} & 1 yr  & 850-2000 \\
Toohey et al. \cite{Toohey2017} & & 850-1850 \\ \hline
\multicolumn{3}{l}{\textbf{Total solar irradiance }}  \\ 											Delaygue et al.\footnotemark[1]\footnote{when multiple versions were provided by PMIP, the “with-background”-version was considered here} \cite{Delaygue2011} & 1 yr  & 850-1850 \\
Muscheler et al.\footnotemark[1]\footnotemark[2] \cite{Muscheler2007} & 1 yr  & 850-1850 \\
Steinhilber et al.\footnotemark[1] \cite{Steinhilber2009a} & 1 yr  & 850-1850 \\
Vieira et al.\footnotemark[1] \cite{Vieira2010, Krivova2007} & 1 yr & 850-1850 \\
Wang et al.\footnotemark[1]\footnotemark[2] \cite{Wang2005} & 1 yr & 1610-2009 \\
Fröhlich et al. \cite{Frohlich2006} & 36 hr & 1978-2017 \\ \hline
\multicolumn{3}{l}{\textbf{CO$_2$}} \\
Schmidt et al.\footnotemark[1] \cite{Schmidt2012} & 1 yr   & 850-2000 \\ 
Keeling et al. \cite{Keeling1976} & 1 hr & 1970-2016  \\ \hline
 \multicolumn{3}{l}{\textbf{Insolation at 65$^{\circ }$N}} \\
Berger\footnotemark[1]\footnote{computed with the \textbf{\textsf{R}}-package ``Palinsol''  \cite{Crucifix2016}} \cite{Berger1978} & 1 yr, 3 hr & 0-2000 \\ \hline
\end{tabular}
\end{ruledtabular} 
\end{table}

\begin{turnpage}
\begingroup
\squeezetable
\begin{table}[!htbp]
\caption[\,Key specification of paleoclimate data]{\label{tab:proxydata} Key specification of proxy records used to estimate local temperature variability. The first six columns (``ID''-``Proxy'') are taken from the PAGES2k database \cite{Neukom2019}. The ID is additionally marked with a start ($\star$) in case the proxy record was used for calculating the agreement in scaling behavior between models and data. The cutoff identifies whether the proxy record was used for analyzing the potential impact of the recent global warming trend (\autoref{fig:warming_control} and \autoref{fig:beta_diff_supp}). It indicates whether the necessary criteria for time series selection were fulfilled before 1850 (PI) and / or over the full historical (hist) period. Similarly, the selection column gives criteria for calculating the mean of local spectra and their comparison (\autoref{fig:warming_control}). The table spans multiple pages.}
\begin{ruledtabular}
\begin{tabular}{lldddllll}
ID   & Name                                            & Lat   & Lon    & Elev\_masl & Archive         & Proxy                   & cutoff  & selection     \\ \hline
2    & Africa-LakeTanganyi.Tierney.2010                & -6    & 28.5   & 28         & lake sediment   & TEX86                   & PI/hist & loose         \\
3    & Africa-Malawi.Powers.2011                       & -10   & 34.3   & 34         & lake sediment   & TEX86                   & PI/hist & loose         \\
4    & Africa-P178-15.Tierney.2015                     & 12    & 44.3   & 44         & marine sediment & TEX86                   & PI/hist & strong/loose  \\
39   & Arc-BrayaSo.DAndrea.2011                        & 67    & -50.7  & -51        & lake sediment   & alkenone                & PI/hist & loose         \\
40   & Arc-Clegg2010                                   & 61.4  & -143.6 & -144       & lake sediment   & midge                   & PI/hist & loose         \\
49*  & Arc-GulfofAlaska.Wiles.2014                     & 61    & -146.6 & -147       & tree            & TRW                     & PI/hist & strong/loose  \\
50   & Arc-HalletLake.McKay.2008                       & 61.5  & -146.2 & -146       & lake sediment   & BSi                     & PI/hist & strong/loose  \\
53   & Arc-Iceland.Bergthorsson.1969                   & 64.8  & -18.4  & -18        & documents       & historic                & PI/hist & loose         \\
56   & Arc-Kongressvatn.DAndrea.2012                   & 78    & 13.9   & 14         & lake sediment   & alkenone                & PI/hist & loose         \\
57   & Arc-Lake4.Rolland.2009                          & 65.1  & -83.8  & -84        & lake sediment   & chironomid              & PI/hist & loose         \\
59   & Arc-LakeE.DAndrea.2011                          & 67    & -50.7  & -51        & lake sediment   & alkenone                & PI/hist & loose         \\
60   & Arc-LakePieni-Ka.Luoto.2010                     & 64.3  & 30.1   & 30         & lake sediment   & chironomid              & PI/hist & loose         \\
65   & Arc-Luoto2009                                   & 60.3  & 25.4   & 25         & lake sediment   & midge                   & PI/hist & strong/loose  \\
66   & Arc-MackenzieDelta.Porter.2013                  & 68.6  & -133.9 & -134       & tree            & \textit{missing}        & hist    & strong/loose  \\
69   & Arc-MD992275.Jiang.2005                         & 66.5  & -17.7  & -18        & marine sediment & diatom                  & PI/hist & strong/loose  \\
81   & Arc-SoperLakeBaf.Hughen.2000                    & 62.9  & -69.9  & -70        & lake sediment   & varve thickness         & PI/hist & strong/loose  \\
82*  & Arc-StoreggaSlid.Sejrup.2011                    & 63.8  & 5.3    & 5          & marine sediment & foram d18O              & PI/hist & strong/loose  \\
83*  & Arc-Thomas2008                                  & 69.9  & -68.8  & -69        & lake sediment   & varve thickness         & PI/hist & strong/loose  \\
85*  & Arc-Tornetrask.Melvin.2013                      & 68.3  & 19.6   & 20         & tree            & TRW                     & PI/hist & strong/loose  \\
88*  & Arc-Yamalia.Briffa.2013                         & 66.8  & 68     & 68         & tree            & TRW                     & PI/hist & strong/loose  \\
122  & Asia-CentralChina.Wang.1998                     & 29    & 113    & 113        & documents       & historic                & PI/hist & strong/loose  \\
130  & Asia-Chu.2012.Sihailongwanlake                  & 42.2  & 126.4  & 126        & lake sediment   & alkenone                & PI/hist & strong/loose  \\
151  & Asia-EastChina.Wang.1990                        & 30    & 117.5  & 118        & documents       & historic                & PI/hist & strong/loose  \\
152* & Asia-EastChinareg.Wang.1998                     & 34    & 120    & 120        & documents       & historic                & PI/hist & strong/loose  \\
157  & Asia-FujianandTai.Wang.1998                     & 24    & 121    & 121        & documents       & historic                & PI/hist & strong/loose  \\
167  & Asia-Guangdong.Zheng.1982                       & 23.2  & 113.2  & 113        & documents       & historic                & PI/hist & strong/loose  \\
168  & Asia-Guangdongand.Zhang.1980                    & 23.5  & 112.5  & 112        & documents       & historic                & PI/hist & strong/loose  \\
182  & Asia-HunanJiangsu.Zhang.1980                    & 28    & 116.5  & 116        & documents       & historic                & PI/hist & strong/loose  \\
205  & Asia-KunashirIsla.Demezhko.2009                 & 44    & 145.7  & 146        & hybrid          & hybrid                  & PI/hist & strong/loose  \\
212  & Asia-Lowerreaches.Zhang.1980                    & 32.1  & 118.8  & 119        & documents       & historic                & PI/hist & strong/loose  \\
220  & Asia-Middlereache.Zhang.1980                    & 30.5  & 114.5  & 114        & documents       & historic                & PI/hist & strong/loose  \\
273* & Asia-SO9039KGSO1.Munz.2015                      & 24.8  & 65.9   & 66         & marine sediment & foraminifera            & PI/hist & strong/loose  \\
275  & Asia-SourthandMid.Demezhko.2007                 & 55    & 59.5   & 60         & borehole        & borehole                & PI/hist & loose         
\end{tabular}
\end{ruledtabular}
\end{table}
\endgroup
\end{turnpage}

\begin{turnpage}
\begingroup
\squeezetable
\begin{table}
\begin{ruledtabular}
\begin{tabular}{lldddllll}
ID   & Name                                            & Lat   & Lon    & Elev\_masl & Archive         & Proxy                   & cutoff  & selection     \\ \hline
276  & Asia-SouthChina.Wang.1998                       & 23    & 114    & 114        & documents       & historic                & PI/hist & strong/loose  \\
331  & Asia-ZhejiangandF.Zhang.1980                    & 25    & 118    & 118        & documents       & historic                & PI/hist & strong/loose  \\ 
342* & Aus-MtRead.Cook.2006                            & -41.8 & 145.5  & 146        & tree            & TRW                     & PI/hist & strong/loose  \\ 
343* & Aus-Oroko.Cook.2002                             & -43.2 & 170.3  & 170        & tree            & TRW                     & PI/hist & strong/loose  \\
347  & Eur-CentralandEa.Pla.2005                       & 42.5  & 0.8    & 1          & lake sediment   & chrysophyte             & PI/hist & loose         \\
349  & Eur-CentralEu.Dobrovolny.2010                   & 49    & 13     & 13         & documents       & Documentary             & PI/hist & strong/loose  \\
350* & Eur-CoastofPortu.Abrantes.2011                  & 41.1  & -8.9   & -9         & marine sediment & alkenone                & PI/hist & strong/loose  \\
352* & Eur-FinnishLakes.Helama.2014                    & 62    & 28.3   & 28         & tree            & MXD                     & PI/hist & strong/loose  \\
354  & Eur-LakeSilvapla.Larocque-Tobler.2010           & 46.5  & 9.8    & 10         & lake sediment   & chironomid              & PI/hist & strong/loose  \\
355* & Eur-LakeSilvapla.Trachsel.2010                  & 46.5  & 9.8    & 10         & lake sediment   & reflectance             & PI/hist & strong/loose  \\
357* & Eur-NorthIceland.Ran.2011                       & 66.5  & -17.7  & -18        & marine sediment & diatom                  & PI/hist & strong/loose  \\
359* & Eur-Seebergsee.Larocque-Tobler.2012             & 46.1  & 7.5    & 8          & lake sediment   & midge                   & PI/hist & strong/loose  \\
364  & Eur-Stockholm.Leijonhufvud.2010                 & 59.3  & 18.1   & 18         & documents       & historic                & PI/hist & strong/loose  \\
365  & Eur-Tallinn.Tarand.2001                         & 59.4  & 24.8   & 25         & documents       & historic                & PI/hist & strong/loose  \\
391  & NAm-BasinP                                      & 44.5  & -70.1  & -70        & lake sediment   & pollen                  & PI/hist & loose         \\
450  & NAm-ClearP                                      & 33.8  & -79    & -79        & lake sediment   & pollen                  & PI/hist & loose         \\
457  & NAm-ConroyL                                     & 46.3  & -67.9  & -68        & lake sediment   & pollen                  & PI/hist & loose         \\
459  & NAm-DarkL                                       & 45.3  & -91.5  & -92        & lake sediment   & pollen                  & PI/hist & loose         \\
463  & NAm-HellKt                                      & 46.2  & -89.7  & -90        & lake sediment   & pollen                  & PI/hist & loose         \\
466* & NAm-LakeMina                                    & 45.9  & -95.5  & -96        & lake sediment   & pollen                  & PI/hist & strong/loose  \\
467  & NAm-LClouds                                     & 48    & -91    & -91        & lake sediment   & pollen                  & PI/hist & strong/loose  \\
468  & NAm-LittlePineL                                 & 45.3  & -91.5  & -92        & lake sediment   & pollen                  & PI/hist & loose         \\
469* & NAm-LNoir                                       & 45.8  & -75.1  & -75        & lake sediment   & pollen                  & PI/hist & strong/loose  \\
485  & NAm-RubyL                                       & 45.3  & -91.5  & -92        & lake sediment   & pollen                  & PI/hist & loose         \\
507  & O2kLR-AlboranSea-TTR17-1384B.Nieto-Moreno.2012  & 36    & -4.7   & -5         & marine sediment & alkenone                & PI/hist & loose         \\
508  & O2kLR-AlboranSea-TTR17-1436B.Nieto-Moreno.2012  & 36.2  & -4.3   & -4         & marine sediment & alkenone                & PI/hist & loose         \\
510  & O2kLR-ArabianSea.Doose-Rolinski.2001            & 24.8  & 65.9   & 66         & marine sediment & alkenone                & PI/hist & loose         \\
511  & O2kLR-CapeGhir.Kim.2007                         & 30.9  & -10.3  & -10        & marine sediment & alkenone                & PI/hist & loose         \\
512  & O2kLR-CapeGhir.McGregor.2007                    & 30.8  & -10.1  & -10        & marine sediment & alkenone                & PI/hist & loose         \\
513  & O2kLR-CapeHatteras.Clroux.2012                  & 35    & -75.2  & -75        & marine sediment & foram Mg/Ca             & PI/hist & strong/loose  \\
514* & O2kLR-CariacoBasin.Black.2007                   & 10.8  & -64.8  & -65        & marine sediment & foram Mg/Ca             & PI/hist & strong/loose  \\
517  & O2kLR-ChileanMargin.Lamy.2002                   & -41   & -74.5  & -74        & marine sediment & alkenone                & PI/hist & loose         \\
518  & O2kLR-DryTortugas.Lund.2006                     & 24.6  & -83.6  & -84        & marine sediment & foram Mg/Ca             & PI/hist & loose         \\
\end{tabular}
\end{ruledtabular}
\end{table}
\endgroup
\end{turnpage}

\begin{turnpage}
\begingroup
\squeezetable
\begin{table}
\begin{ruledtabular}
\begin{tabular}{lldddllll}
ID   & Name                                            & Lat   & Lon    & Elev\_masl & Archive         & Proxy                   & cutoff  & selection     \\ \hline
519  & O2kLR-DryTortugasA.Lund.2006                    & 24.3  & -83.3  & -83        & marine sediment & foram Mg/Ca             & PI/hist & loose         \\
520  & O2kLR-EasternTropicalNorthAtlantic.Kuhnert.2011 & 16.8  & -16.7  & -17        & marine sediment & foram Mg/Ca             & PI/hist & loose         \\
521  & O2kLR-EmeraldBasin.Keigwin.2003                 & 45.9  & -62.8  & -63        & marine sediment & alkenone                & PI/hist & loose         \\
523  & O2kLR-FeniDriftRichter.2009                     & 55.5  & -13.9  & -14        & marine sediment & foram Mg/Ca             & PI/hist & loose         \\
524  & O2kLR-FiskBasin.Richey.2009                     & 27.6  & -93.9  & -94        & marine sediment & foram Mg/Ca             & PI/hist & strong/loose  \\
525  & O2kLR-GarrisonBasin.Richey.2009                 & 26.7  & -93.9  & -94        & marine sediment & foram Mg/Ca             & hist    & loose         \\
526  & O2kLR-GreatBahamaBank.Lund.2006                 & 24.6  & -79.3  & -79        & marine sediment & foram Mg/Ca             & PI/hist & loose         \\
527  & O2kLR-GreatBahamabAtlantic0326aLund2006125MC    & 24.8  & -79.3  & -79        & marine sediment & foram Mg/Ca             & PI/hist & loose         \\
528  & O2kLR-GreatBarrier.Hendy.2002                   & -18.3 & 146.6  & 147        & coral           & Coral Sr/Ca             & PI/hist & strong/loose  \\
529  & O2kLR-GulfofGuinea.Weldeab.2007                 & 2.5   & 9.4    & 9          & marine sediment & foram Mg/Ca             & PI/hist & loose         \\
530  & O2kLR-JacafFjord.Sepulveda.2009                 & -44.3 & -73    & -73        & marine sediment & alkenone                & PI/hist & loose         \\
532  & O2kLR-KuroshioCurrent.Isono.2009                & 36    & 141.8  & 142        & marine sediment & alkenone                & PI/hist & loose         \\
535  & O2kLR-MakassarStrait-MD98-2177.Newton.2011      & 1.4   & 119.1  & 119        & marine sediment & foram Mg/Ca             & PI/hist & strong/loose  \\
536  & O2kLR-MakassarStrait.Linsley.2010               & -7.4  & 115.2  & 115        & marine sediment & foram Mg/Ca             & PI/hist & loose         \\
538  & O2kLR-MakassarStrait.Oppo.2009                  & -3.5  & 119.2  & 119        & marine sediment & foram Mg/Ca             & PI/hist & loose         \\
539  & O2kLR-MD952011.Calvo.2002                       & 67    & 7.6    & 8          & marine sediment & alkenone                & PI/hist & loose         \\ 
540  & O2kLR-Minorca.Moreno.2012                       & 40.5  & 4      & 4          & marine sediment & alkenone                & PI/hist & loose         \\
541* & O2kLR-NorthIceland.Sicre.2011                   & 66.5  & -17.4  & -17        & marine sediment & alkenone                & PI/hist & strong/loose  \\
542  & O2kLR-NWPacific.Harada.2004                     & 46.3  & 152.5  & 152        & marine sediment & alkenone                & PI/hist & loose         \\
544  & O2kLR-OkinawaTrough.Wu.2012                     & 24.8  & 122.5  & 122        & marine sediment & TEX86                   & PI/hist & loose         \\
546  & O2kLR-Philippines-MD98-2181.Stott.2007          & 6.3   & 125.8  & 126        & marine sediment & foram Mg/Ca             & PI/hist & loose         \\
548  & O2kLR-PigmyBasin.Richey.2009                    & 27.2  & -91.4  & -91        & marine sediment & foram Mg/Ca             & PI/hist & strong/loose  \\
550* & O2kLR-SantaBarbara.Zhao.2000                    & 34.2  & -120   & -120       & marine sediment & alkenone                & PI/hist & strong/loose  \\
551  & O2kLR-SouthAtlantic.Leduc.2010                  & -29.1 & 16.7   & 17         & marine sediment & alkenone                & PI/hist & loose         \\
552  & O2kLR-SouthChinaSea.Zhao.2006                   & 8.7   & 109.9  & 110        & marine sediment & alkenone                & PI/hist & loose         \\
553  & O2kLR-SouthernChile.Mohtadi.2007                & -44.1 & -75.2  & -75        & marine sediment & alkenone                & PI/hist & loose         \\
554* & O2kLR-SouthIceland.Sicre.2011                   & 57.5  & -27.9  & -28        & marine sediment & alkenone                & PI/hist & strong/loose  \\
555  & O2kLR-SubTropicalEasternAtlantic.deMenocal.2000 & 20.8  & -18.6  & -19        & marine sediment & planktonic foraminifera & PI/hist & loose         \\
561  & O2kLR-WesternSvalbard.Spielhagen.2011           & 78.9  & 6.8    & 7          & marine sediment & planktonic foraminifera & PI/hist & loose         \\
562  & O2kLR-WestSpitzbergen.Bonnet.2010               & 79    & 5.9    & 6          & marine sediment & dynocist MAT            & PI/hist & loose         \\
563  & O2kLRMakassarStrait-MD98-2160.Newton.2011       & -5.2  & 117.5  & 118        & marine sediment & foram Mg/Ca             & PI/hist & loose         \\
654* & SAm-LagunaAculeo.vonGunten.2009                 & -33.9 & -70.9  & -71        & lake sediment   & reflectance             & PI/hist & strong/loose  \\
655* & SAm-LagunaChepical.deJong.2013                  & -32.3 & -70.5  & -70        & lake sediment   & reflectance             & PI/hist & strong/loose  \\
656* & SAm-LagunaEscondida.Elbert.2013                 & -45.5 & -71.8  & -72        & lake sediment   & BSi                     & PI/hist & strong/loose 

\end{tabular}
\end{ruledtabular}
\end{table}
\endgroup
\end{turnpage}

\clearpage

\begin{figure*}[!htbp]
\includegraphics[width=17.2cm]{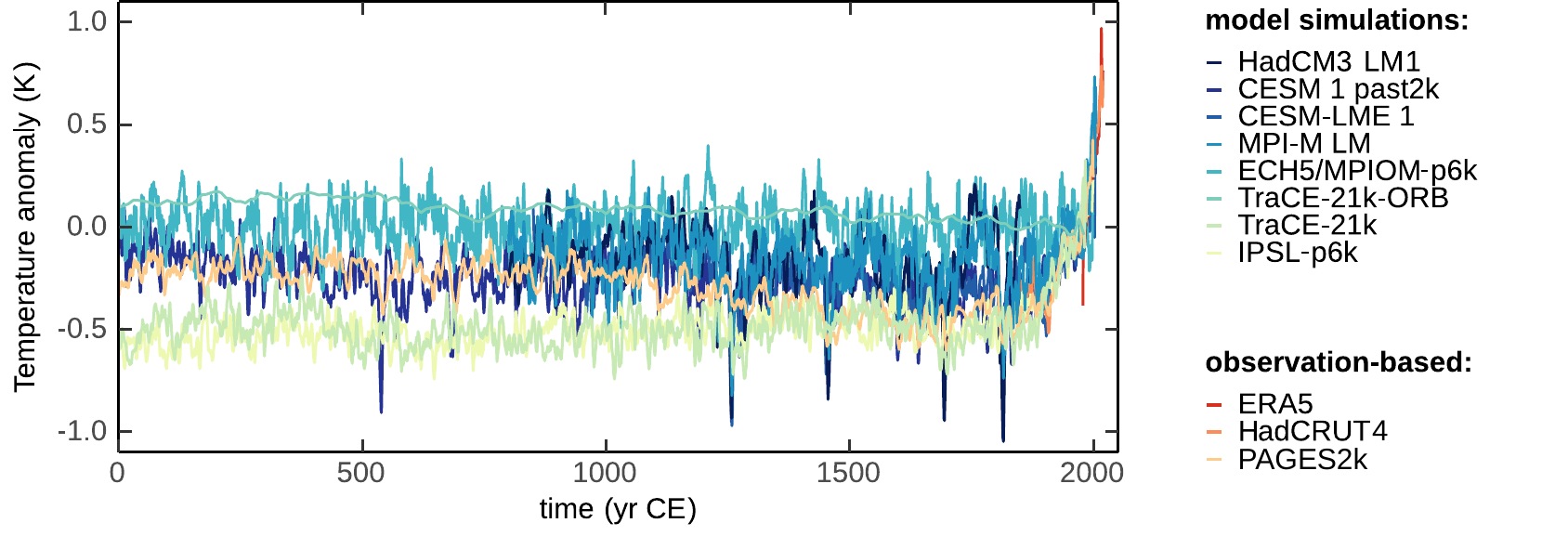}%
\caption[\,Global mean temperature over the Common Era]{\label{fig:models_supp} Evolution of global mean temperature over the Common Era from model simulations and observation-based data as in \autoref{tab:signal_data}. Anomalies are given with respect to the reference period 1961-1990 (HadCM3 LM1:1800-1850) and as a running average of five years.}
\end{figure*}

\begin{figure}[!htbp]
\includegraphics[width=17.2cm]{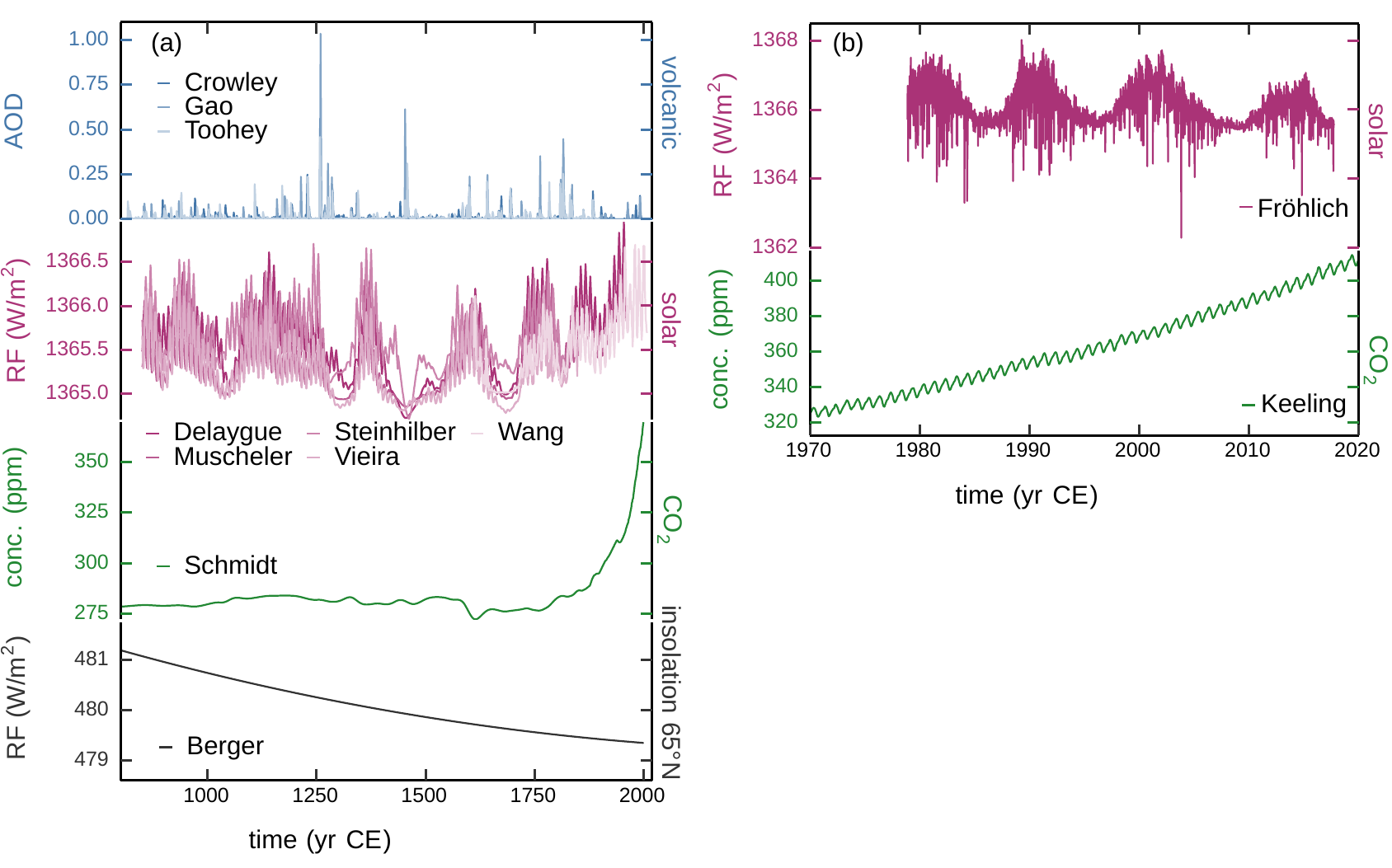}%
\caption[\,Climatic drivers over the last millennium]{\label{fig:forcings_supp} (a) Reconstruction of climate drivers over the Common Era used to estimate the PSD of radiative forcing (RF). Labels indicate the data reference as given in \autoref{tab:forcing_data}. (b) Additional observational data for solar and CO$_2$ forcing used to obtain high-frequency spectral estimates. Highly resolved insolation changes due to the diurnal and annual cycle are not shown here.}
\end{figure}

\begin{figure}[!htbp]
\includegraphics[width=17.2cm]{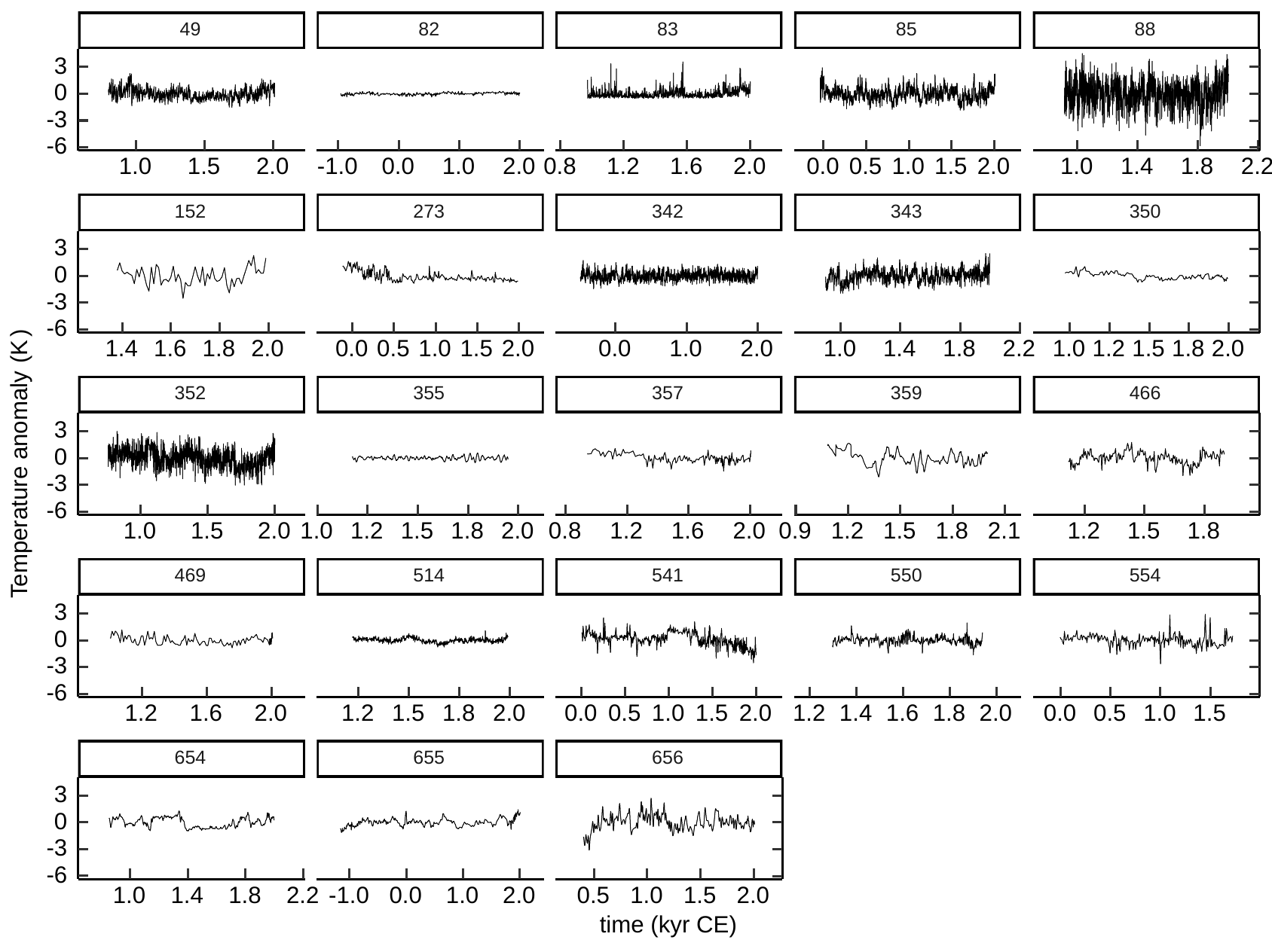}%
\caption[\,Temperature signals from proxy records over the last millennia]{\label{fig:proxies_supp} Temperature anomalies from proxy records used to estimate local temperature variability. The labels give the ID from the PAGES2k database as indicated by a star ($\star$) in \autoref{tab:proxydata}.}
\end{figure}

\begin{figure*}[!htbp]
\includegraphics[width=17.2cm]{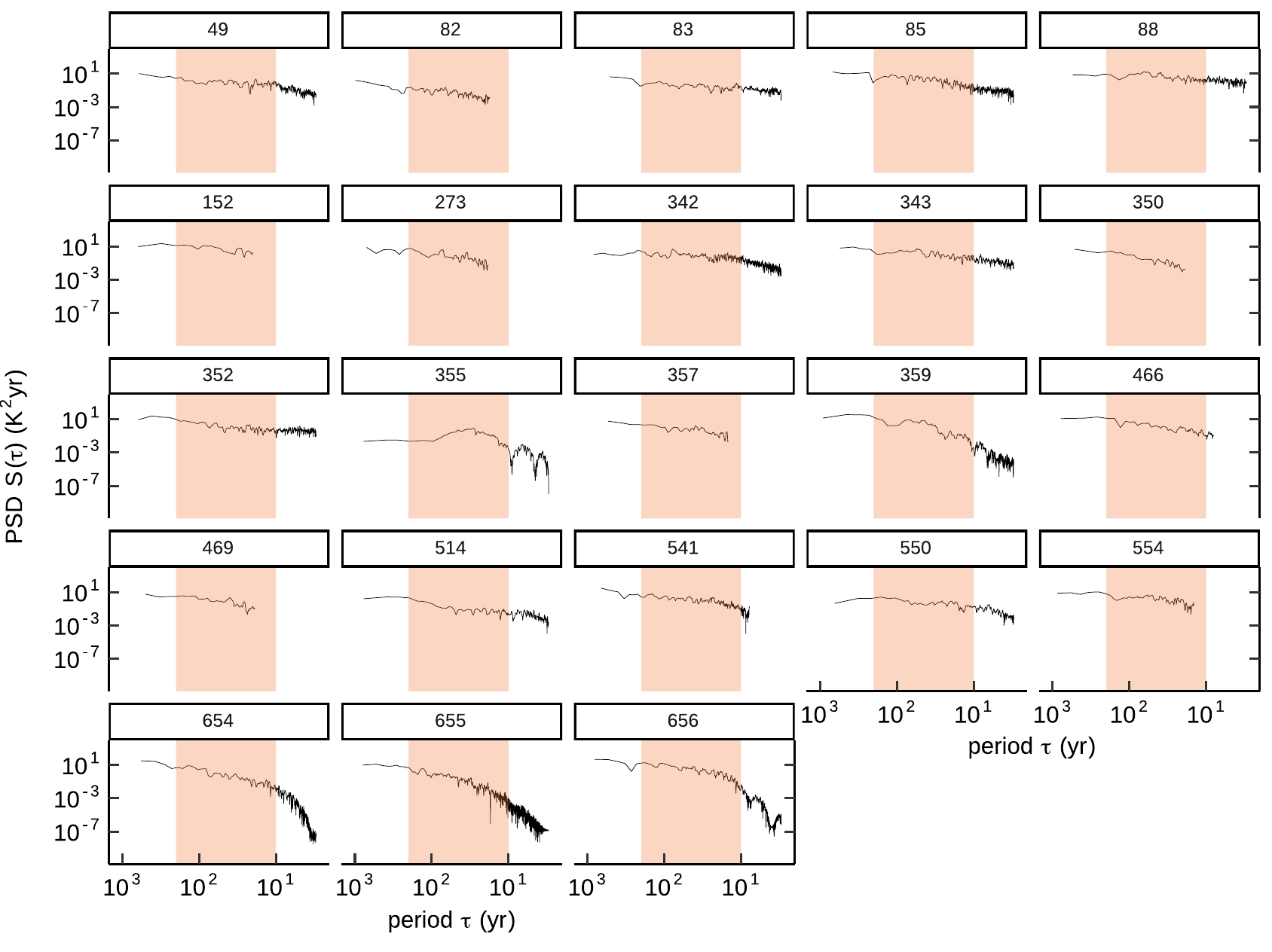}%
\caption[\,PSD of proxy records]{\label{fig:spectra_supp}As \autoref{fig:proxies_supp}, but showing the PSD of the proxy records. The shaded area indicates the periods between 10 and 200 years, used to estimate the scaling relationship. Scaling exponents for spectra that do not cover the full period were estimated on their corresponding timescales, but at least between 20 and 200 years.}
\end{figure*}

\begin{figure*}[!htbp]
\includegraphics[width=17.2cm]{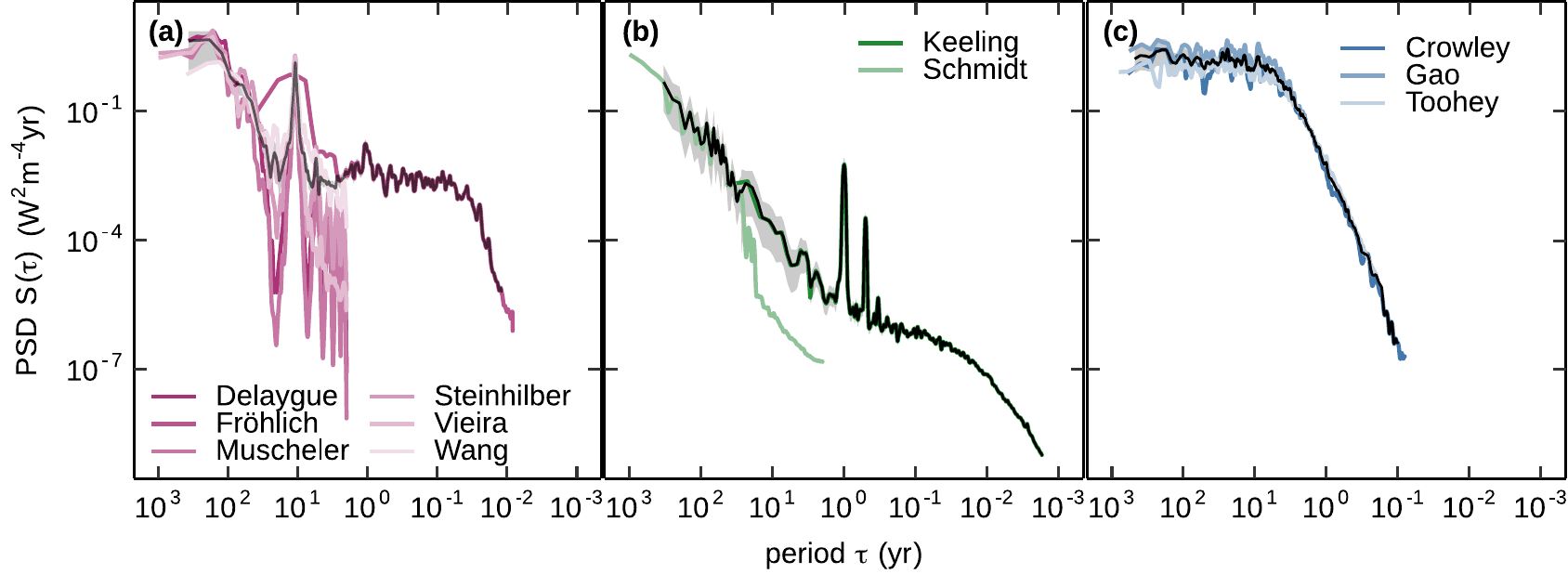}%
\caption[\,Mean PSDs for radiative forcing reconstructions]{\label{fig:forcing_composite} Power spectral densities of solar (a), CO$_2$ (b), and volcanic (c) from multiple data sets and their mean spectra (black). Low- and high-frequency tails, likely prone to spectral biases, were omitted in the average.}
\end{figure*}

\begin{figure}[!htbp]
\includegraphics[width=8.6cm]{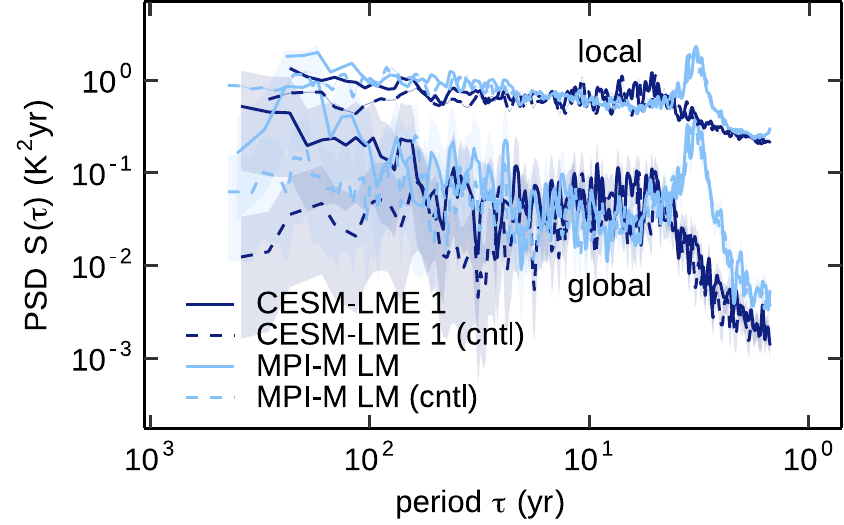}%
\caption[PSD from last millennium runs and their PI control runs]{\label{fig:warming_control} PSD for global mean and mean of local spectra of surface air temperature from last millennium runs and their PI control (cntl).}
\end{figure}

\begin{figure}[!htbp]
\includegraphics[width=8.6cm]{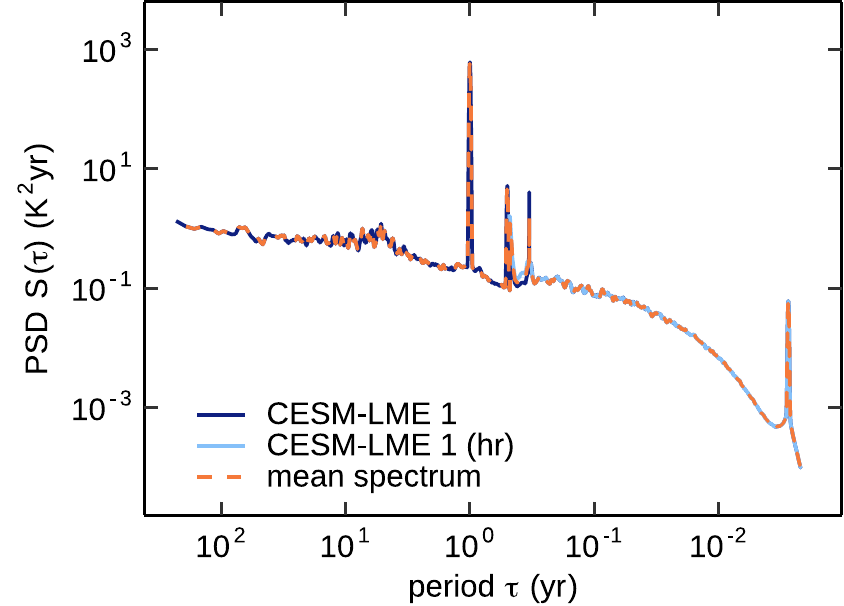}%
\caption[\,Mean PSD from high- and low-frequency spectrum]{\label{fig:patchwork_supp}
Joint PSD (dashed orange line) from the mean spectrum of a highly (hr) and lower resolved time series on the example of surface air temperature from the CESM-LME 1 run. To improve computational efficiency, we formed this joint PSD for ERA5, MPI-M LM, and CESM-LME 1. The high-frequency component was estimated from data at hourly resolved temperature series from 1981 to 1990 CE since this is a part of the climatic period that is shared among the three data sets.}
\end{figure}

\begin{figure}[!htbp]
\includegraphics[width=8.6cm]{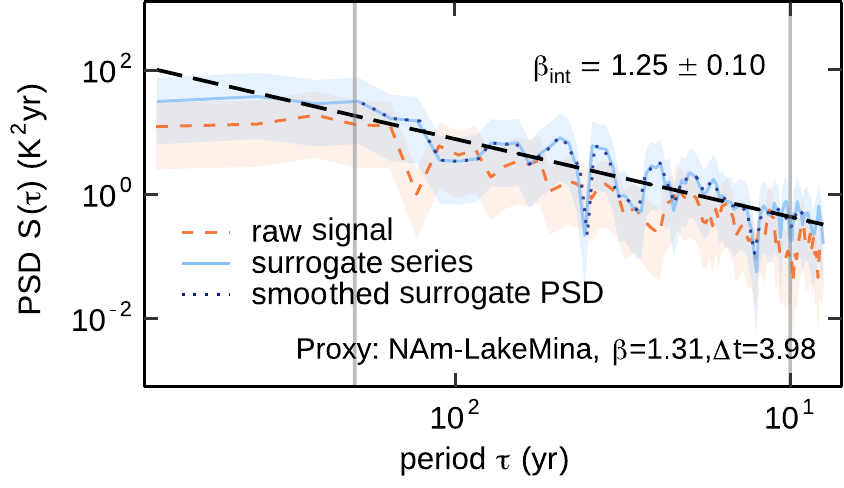}%
\caption[\,Uncertainty quantification for irregularly sampled temperature proxies]{\label{fig:blockavg_supp} Step-wise estimation of uncertainty on $\beta$-scaling relation for irregularly sampled proxy records on the example of the PSD from the ``NAm-LakeMina''-record. When interpolated to its mean temporal resolution $\Delta t$, the raw signal has a scaling relation $\beta=1.31$ (orange dashed line). The uncertainties were calculated from 100 surrogates that are random time series with approximately the same power-law scaling $\beta$ (black dotted line). The PSD is altered after forming the block-average of the surrogate time series and interpolation to $\Delta t$ (blue solid line). Linear regression on this spectrum (black dashed line) gives the scaling coefficient $\beta_{\mathrm{int}}$ of the surrogate series.}
\end{figure}

\begin{figure}[!htbp]
\includegraphics[width=8.6cm]{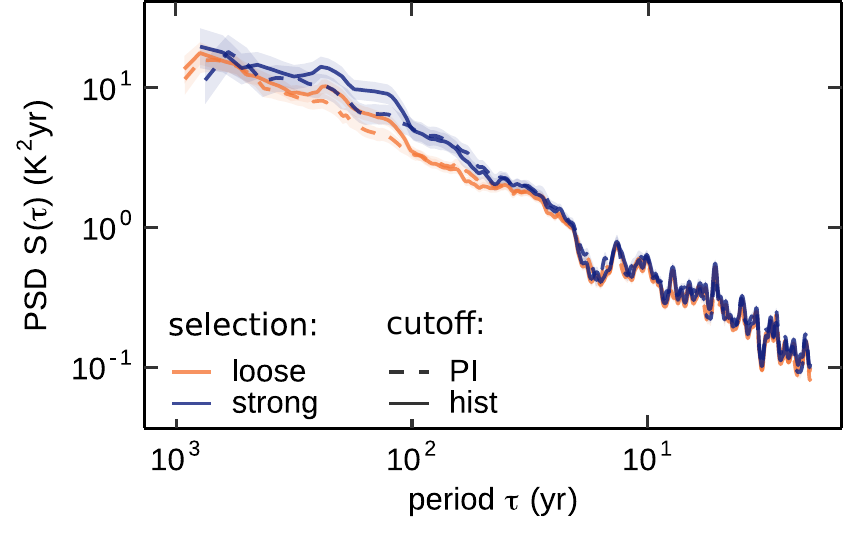}%
\caption[\,The effect of sampling on the mean of local PSDs from proxy records]{\label{fig:pages_comp_supp}Global mean of local PSD from proxy records subject to sampling from the PAGES2k data base and global warming. Dashed and solid lines refer to the mean spectra of local proxy records cut at 1850 CE (PI) and 2000 CE (hist) respectively. The plot compares strong (blue) and loose (orange) criteria. Loose criteria are described in the data section of the main manuscript and were used to calculate the mean of local spectra from reconstructions. For the stronger criteria we require the mean temporal resolution ($\langle t_{i+1}$ - $t_i \rangle) \leq 20 $years, a coverage ($t_N$ - $t_1$) $\geq 30$years and the number of data points $N \geq$ 30. Hiatus are tolerated up to max($t_{i+1}$- $t_i$) $\geq 40$ years.
}
\end{figure}

\begin{figure}[!htbp]
\includegraphics[width=8.6cm]{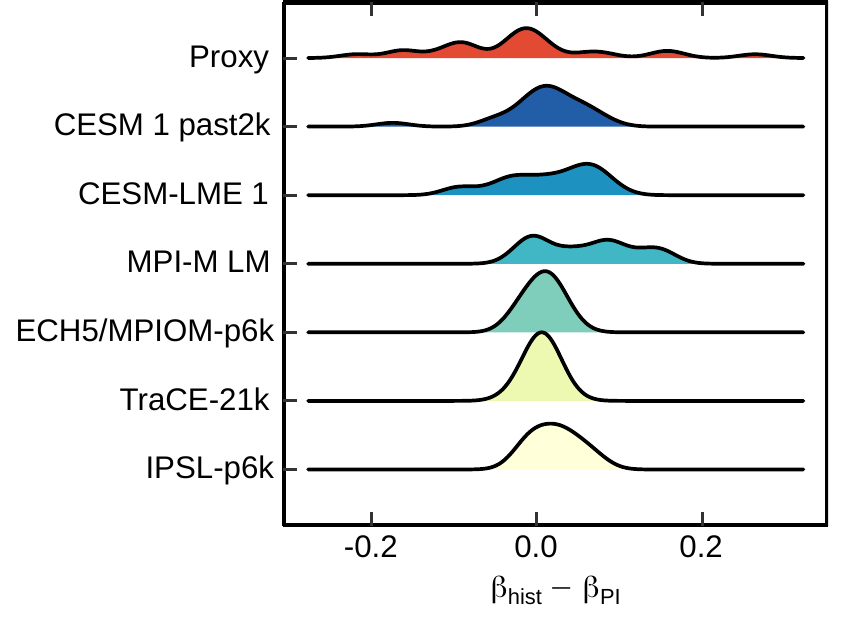}%
\caption[\,The effect of global warming on the spectral exponent $\beta$]{\label{fig:beta_diff_supp} Normalized density plots of the deviations of the spectral exponent $\beta$ extracted from time series up to 2020 CE ($\beta_{\mathrm{hist}}$) and time series up to 1850 CE ($\beta_{\mathrm{PI}}$). The densities were computed from the local power-law scaling used to estimate the statistical agreement between models and data  (\autoref{tab:proxydata}).}
\end{figure}

\begin{figure*}[!htbp]
\includegraphics[width=17.2cm]{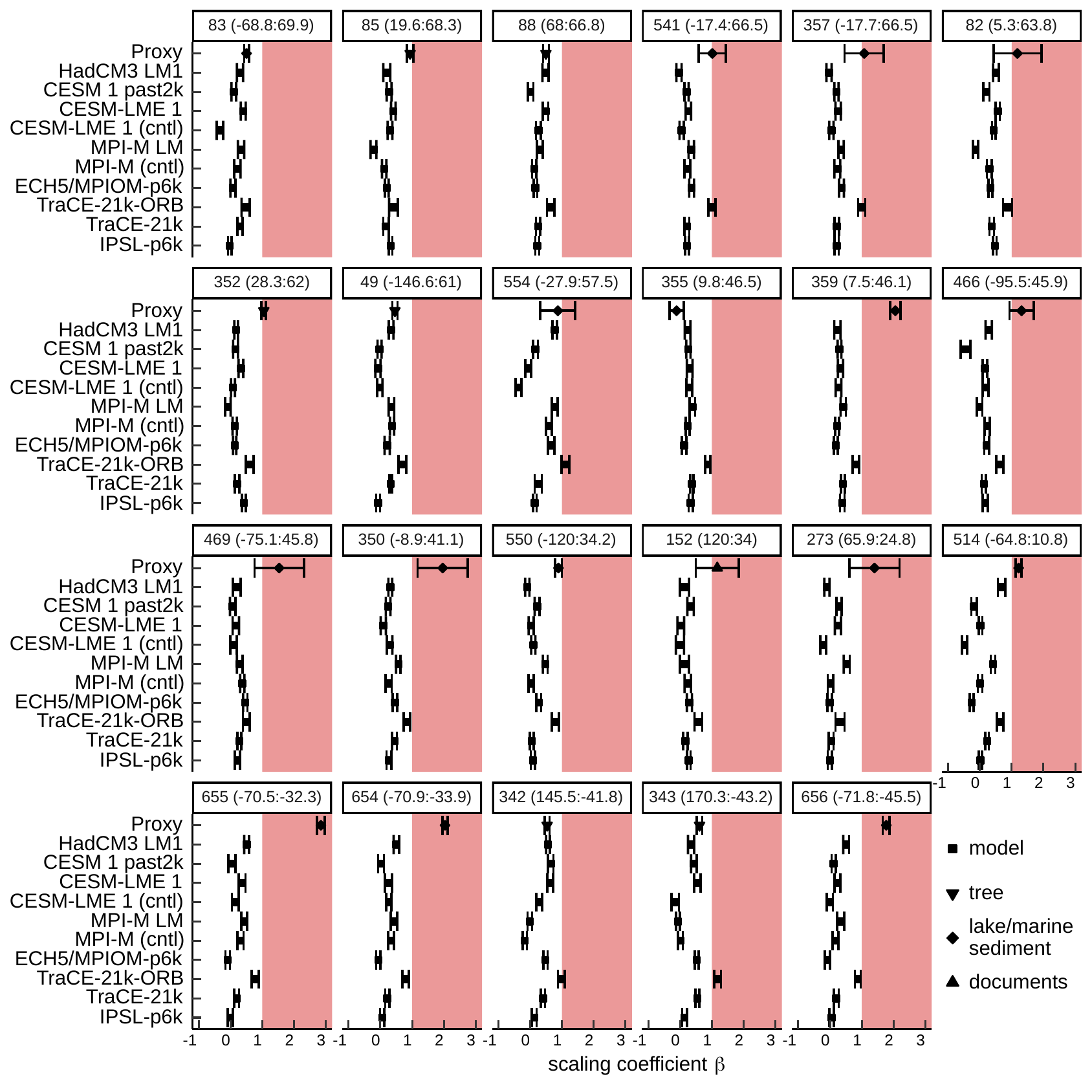}
\caption[\,Scaling of local temperature from simulation and proxies]{\label{fig:statistics_all} Local scaling coefficients and its confidence intervals from simulations and paleoclimate data. Each panel denotes the ID within the PAGES2k database and its coordinates (longitude $^\circ$E, latitude $^\circ$N). Symbols indicate the archive of the proxy record or whether the scaling exponent belongs to a model simulation or not. Red shading refers to scaling exponents $\beta >1$, whereas the white background indicates $\beta <1$.}
\end{figure*}

\begin{figure*}[!htbp]
\includegraphics[width=17.2cm]{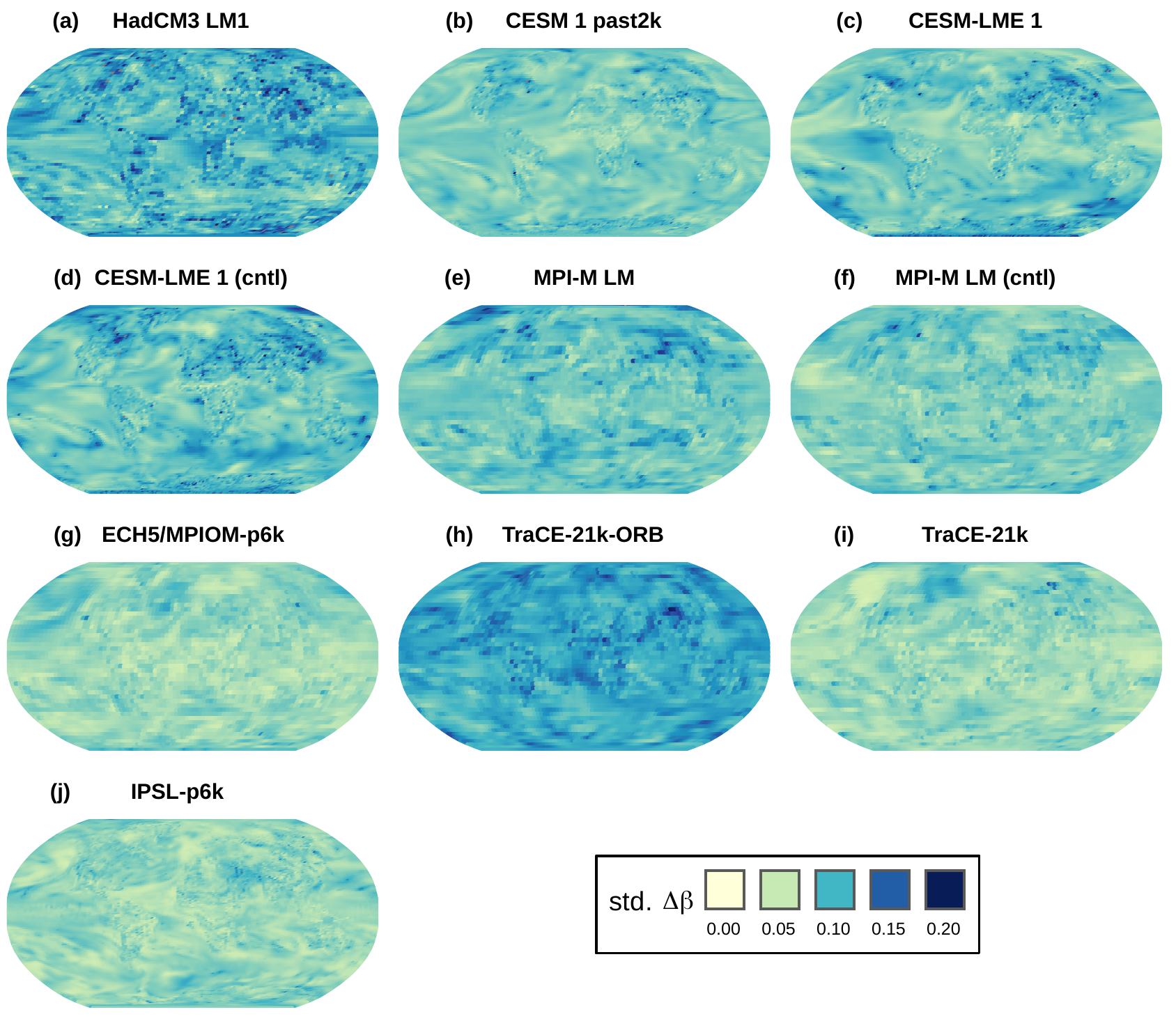}%
\caption[\,Goodness of fit for local scaling coefficients from model simulations]{\label{fig:goodness_of_fit} Standard deviation of the least-square regression of local scaling coefficient $\beta$ for all model simulations. The goodness of fit is generally lower for models with a comparatively low temporal resolution (Trace21k-ORB) or coverage (HadCM3 LM1).  Furthermore, there is a slightly increased deviation in areas of active modes of variability, such as the ENSO region ((b), (c), (e), (j)). This could be due to the fact, that spectral peaks describe these (quasi-)periodic signals better than power-law scaling.}
\end{figure*}

\begin{figure*}[!htbp]
    \centering
    \includegraphics[width=17.2cm]{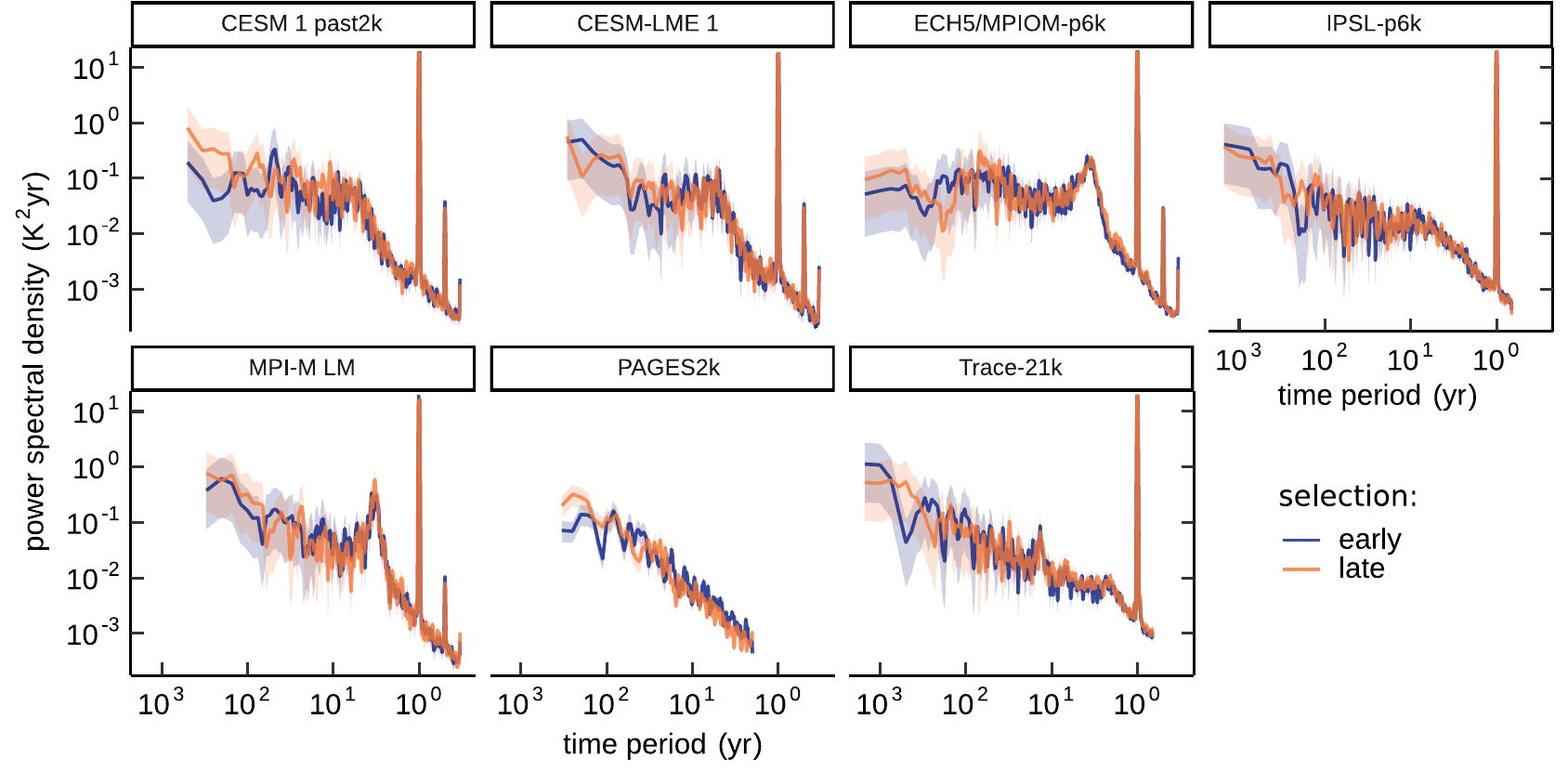}
    \caption[\,PSD of disjoint time series of global mean temperature]{\label{fig:disjoint} PSD of global mean surface temperature from model simulations and PAGES2k reconstruction. To test the robustness of our spectral analysis against potentially non-trending, non-stationary signals in the data sets, each temperature signal was split into two disjoint time intervals of equal length, representing the ``early'' and the ``late'' part of the signal. The latter contains the anthropogenic global warming period, starting approximately 1850 CE. Trace-21k-ORB (orbitally forced only) as well as HadCM3 LM1 (800-1850 CE) were excluded from the discussion since they do not represent the recent global warming.}
\end{figure*}

\begin{figure}[!htbp]
    \centering
    \includegraphics[width=8.6cm]{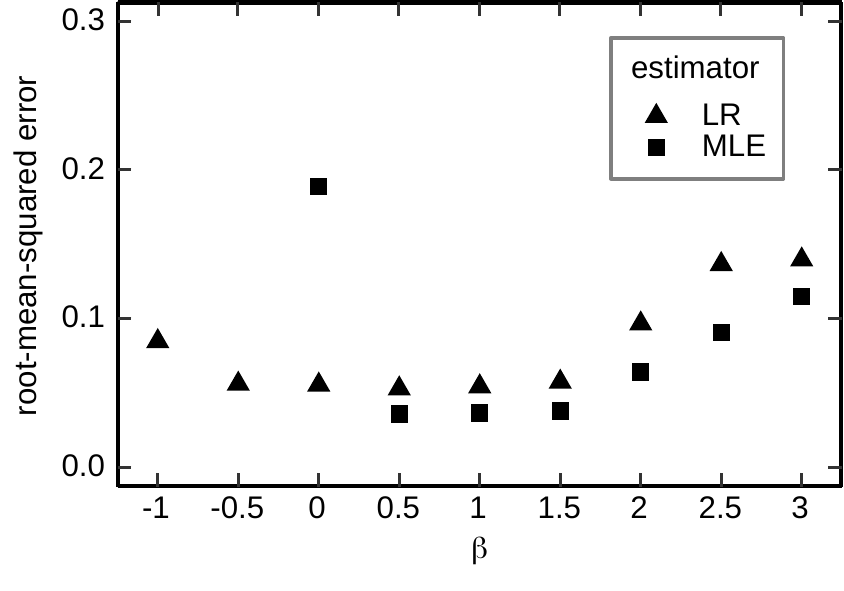}
    \caption[\,Root-mean-squared error of LR and MLE]{\label{fig:MSE} Root-mean-squared error of linear regression (LR) and maximum likelihood estimation (MLE) \cite{Gillespie2020, Clauset2009} computed from 200 surrogate time series with 6000 data points and power-law scaling $\beta$. The grey dashed line marks the mean confidence $\Delta \beta_{\mathrm{LR}}$ of the linear regression. Outliers ($> 0.3$) of the MLE for negative scaling exponents are not shown here. The implemented (standard) MLE is generally not appropriate to estimate $\beta \leq 0$, which explains deviations in this range \cite{Clauset2009}.}
\end{figure}

\begin{figure*}[!htbp]
    \centering
    \includegraphics[width=17.2cm]{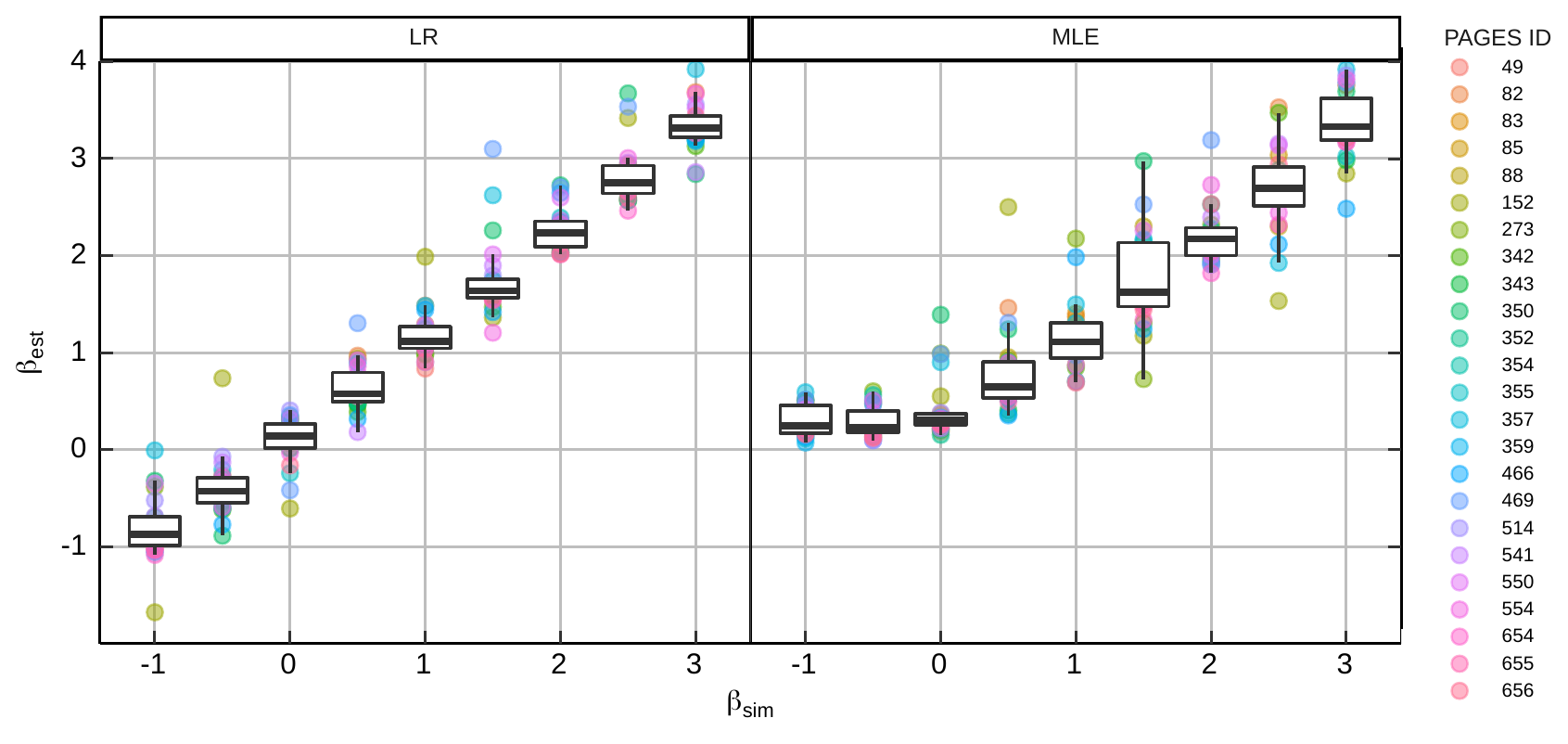}
    \caption[\,Comparison of LR and MLE for irregularly sampled data]{\label{fig:beta_irregular} Distribution of estimated beta ($\beta_{\mathrm{est}}$) from irregularly sampled pseudo proxy records using LR and MLE. For each possible $\beta_{\mathrm{sim}}$, a set of 200 surrogate time series was block-averaged to the resolution of each proxy record before computing its power spectral density. The horizontal thick lines within the boxes correspond to the median. The ends of the box denote the upper (75\%) and lower (25\%) quartiles. The vertical line extends from the upper ending of the box to the largest value no further than 1.5 times the interquartile range, and vice versa for the lower bound. Points further outside are potential outliers, of which the IDs 152, 273, 350, and 469 can be explained by their comparatively low mean resolution of 9-10 years.}
\end{figure*}

\begin{figure}[!htbp]
    \centering
    \includegraphics[width=8.6cm]{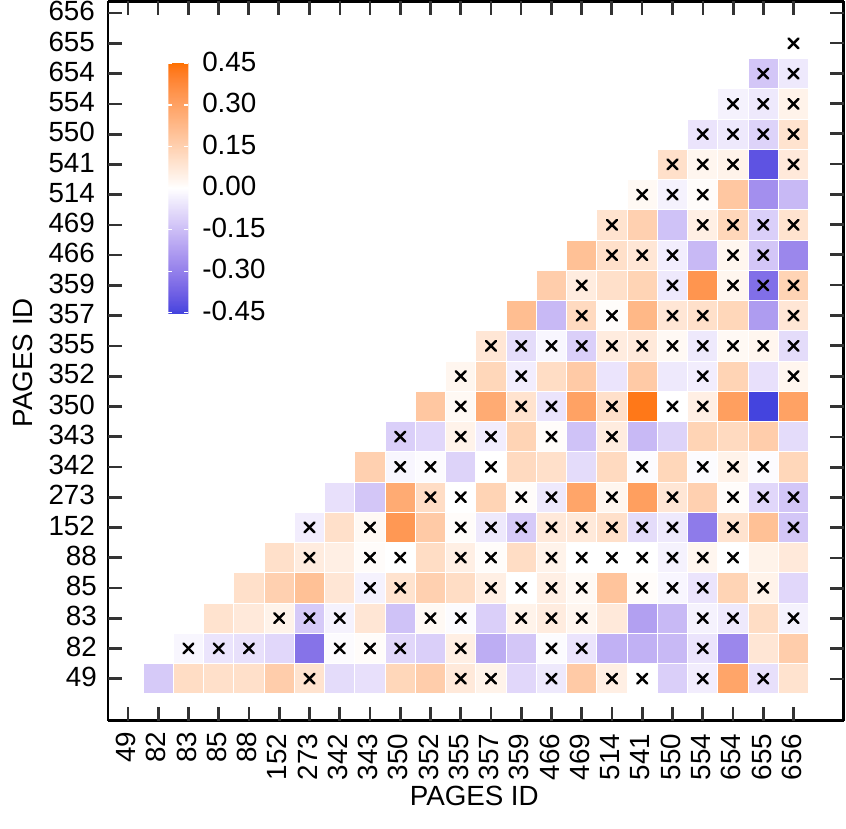}
    \caption[\,Cross-correlations of the examined proxy records]{\label{fig:crosscorrelation} Cross-correlations among the 23 proxy records, denoted by the ID of the PAGES2k database \cite{Neukom2019}. Insignificant correlations are marked by a cross. The cross-correlations for the irregularly sampled data were computed using the \textit{nest}-package \cite{Rehfeld2011, Rehfeld2014} in \textsf{R}.}
\end{figure}

\clearpage
\section*{ }